\documentclass[aps,prx,twocolumn,floatfix,superscriptaddress,nofootinbib]{revtex4}

\usepackage{amssymb,amsmath,amstext}                
\usepackage{graphicx}                                               
\usepackage{epstopdf}                                               
\usepackage{color}                                                     
\usepackage{bm}                                                        
\usepackage{appendix}                                              
\usepackage[utf8]{inputenc}
\usepackage{bbold}
\usepackage{bbm}
\usepackage{ulem}
\usepackage{bm}
\usepackage{latexsym}
\usepackage[colorlinks=true,citecolor=blue,linkcolor=magenta]{hyperref}

\def\be{\begin{equation}}
\def\ee{\end{equation}}
\def\bea{\begin{eqnarray}}
\def\eea{\end{eqnarray}}

\def\bi{\begin{itemize}}
\def\ei{\end{itemize}}
\def\ben{\begin{enumerate}}
\def\een{\end{enumerate}}

\newcommand{\vct}[1]{\mathbf{#1}}

\begin{document} 

\title{Skin effect as a probe of transport regimes in Weyl semimetals}

\author{Pawe\l{} Matus} 
\email{matus@pks.mpg.de}
\affiliation{Max Planck Institute for the Physics of Complex Systems and W\"{u}rzburg-Dresden Cluster of Excellence ct.qmat, 01187 Dresden, Germany}
\author{Renato M. A. Dantas} 
\email{renatomiguel.alvesdantas@unibas.ch }
\affiliation{Max Planck Institute for the Physics of Complex Systems and W\"{u}rzburg-Dresden Cluster of Excellence ct.qmat, 01187 Dresden, Germany}
\affiliation{Department of Physics, University of Basel, Klingelbergstrasse 82, 4056 Basel, Switzerland}
\author{Roderich Moessner} 
\email{moessner@pks.mpg.de}
\affiliation{Max Planck Institute for the Physics of Complex Systems and W\"{u}rzburg-Dresden Cluster of Excellence ct.qmat, 01187 Dresden, Germany}
\author{Piotr Sur\'{o}wka} 
\email{surowka@pks.mpg.de}
\affiliation{Department of Theoretical Physics, Wroc\l{}aw  University  of  Science  and  Technology,  50-370  Wroc\l{}aw,  Poland}
\affiliation{Max Planck Institute for the Physics of Complex Systems and W\"{u}rzburg-Dresden Cluster of Excellence ct.qmat, 01187 Dresden, Germany}

\begin{abstract}
We study the propagation of an oscillatory electromagnetic field inside a Weyl semimetal. In conventional conductors, the motion of the charge carriers in the skin layer near the surface can be diffusive, ballistic, or hydrodynamic. We show that the presence of chiral anomalies, intrinsic to the massless Weyl particles, leads to a hitherto neglected nonlocal regime that can separate the normal and viscous skin effects. We propose to use this novel regime as a diagnostic of the presence of chiral anomalies in optical conductivity measurements. These results are obtained from a generalized kinetic theory which includes various relaxation mechanisms, allowing us to investigate different transport regimes of Weyl semimetals.
\end{abstract}

\date{\today}

\maketitle

\section{Introduction}

Electronic transport phenomena often present particularly direct and accurate probes of the properties of conducting materials. Conventional metals are well described by the Drude theory that captures the diffusive movement of charge carriers. In very clean materials, where the scattering of electrons with impurities is not frequent, transport departs from being diffusive. Depending on the interaction strength, transport exhibits either ballistic or hydrodynamic behavior \cite{Gurzhi1963,Gurzhi1968}. So far the experimental candidates to investigate such effects are two-dimensional (Al,Ga)As heterostructures \cite{Molenkamp1994,deJong1995,Gusev2018}, and graphene \cite{crossno_observation_2016,bandurin_negative_2016,bandurin_fluidity_2018,sulpizio_visualizing_2019,berdyugin_measuring_2019,ku_imaging_2020,keser_geometric_2021}. A detailed theoretical analysis describing viscous electronics and ballistic-to-hydrodynamic cross-overs in clean two-dimensional structures is done by means of kinetic theory with various relaxation times corresponding to different scattering mechanisms \cite{narozhny_hydrodynamic_2017,Lucas2018}.

In a parallel line of developments, much effort has been devoted to the study of topological effects in the transport of massless quasiparticles. In three dimensions, these can be realized as low-energy electronic excitations in materials known as Dirac semimetals and Weyl semimetals. In Weyl semimetals either time reversal symmetry or inversion symmetry is absent and the spectrum in the vicinity of the Fermi energy consists of an even number of linear band crossings known as Weyl nodes. Each crossing hosts fermions with a well-defined chirality, and the sum of the chiralities in the Brillouin zone is zero \cite{Armitage2018, Nielsen1981}. The fact that the Weyl nodes are generically separated in the reciprocal space makes these systems an ideal platform to investigate transport phenomena related to the chiral anomaly. It describes the breaking of the classical chiral symmetry by quantum fluctuations in parallel electric and magnetic fields \cite{Nielsen1983, Bertlmann:1996xk,Landsteiner:2016led}. At weak coupling, the imprint of the chiral anomaly has been connected to the negative magnetoresistance observed in Dirac and Weyl semimetals \cite{huang_observation_2015,li_giant_2015,li_negative_2016,li_chiral_2016,kim_dirac_2013,xiong_evidence_2015,zhang_signatures_2016,liang_experimental_2018,Dantas2018}. In the hydrodynamic regime, macroscopic effects tied to chiral anomalies are the chiral magnetic effect (CME), the chiral vortical effect and thermal transport phenomena related to gravitational anomalies \cite{Fukushima2008,Kharzeev:2009pj,Banerjee:2008th,Erdmenger:2008rm,Neiman:2010zi,Son:2009tf,Landsteiner:2011cp,lucas_hydrodynamic_2016,Gorbar2018,Dantas2020}. Despite the above predictions, on the one hand, the direct identification of the chiral anomalies with negative magnetoresistance is problematic due to other interfering effects \cite{ong_experimental_2021}; on the other hand, the hydrodynamic regime is more difficult to control experimentally (for recent progress see refs. \cite{Gooth2018,vool_imaging_2021}) and the corresponding effects difficult to measure in practice.

The relevant degrees of freedom to capture transport in semimetals with a finite Fermi energy and at low temperatures can be modeled as long-lived quasiparticles. In this case the pertinent quantity is the single particle distribution function due to the fact that we can treat interactions perturbatively. The evolution of the single particle distribution function is given by the Boltzmann kinetic equation, which incorporates both the single particle dynamics and the relaxation processes arising from interactions between constituents of the system. This provides the semiclassical description of physical observables that vary slowly in space and time, establishing a statistical description of a many-body electron system. The effect of collisions of the particles has to be encapsulated by an appropriate choice of the collision term, which is usually done by a phenomenological treatment of the appropriate relaxation mechanisms \cite{Lifshitz,Soto}. In the case of massless fermions under the influence of an external magnetic field, the classical evolution gets modified due to the Berry phase contribution. Thus, the dynamics of the distribution function also changes and is given by the chiral kinetic equation \cite{Loganayagam:2012pz,SonYamamoto2012,Stephanov2012,Chen2014,Chen2015}.

Transport properties of Weyl semimetals are not limited to steady flow configurations. In fact, semimetals driven by alternating current (ac) electric fields exhibit a rich phenomenology and reveal new effects absent in direct current (dc) settings both in two \cite{Yoshikawa2017, moessner_pulsating_2018,semenyakin_alternating_2018, Hafez2018, moessner_boundary-condition_2019,alekseev_magnetic_2018,chandra_hydrodynamic_2019,levchenko_transport_2020, McIver2020} and three dimensions \cite{Sodemann2015, morimoto_semiclassical_2016,burkov_dynamical_2018,Osterhoudt2019,golub_semiclassical_2020,Kovalev2020, Cheng2020, sukhachov_stray_2021,weber_ultrafast_2021,Dantas2021}. 

Here, our goal is to investigate the ramifications of chiral anomalies on transport properties at finite frequency. One of the classic examples of a finite driving transport phenomenon is the skin effect \cite{tanner_optical_2019}. It states that an alternating electric current is distributed mainly close to the wall of the conductor and decays exponentially within the conductor. Therefore, the skin effect increases its effective resistance through the reduction of the effective cross-section of the conductor. This reduction is parameterized by a distance, called the penetration depth or the skin depth, that depends on the driving frequency. Depending on the dominant relaxation mechanism in the skin layer, which in general changes with the skin depth, three types of skin effect are conventionally distinguished: normal, anomalous, and viscous \cite{kaganov_theory_1997,Gurzhi1968}. The normal skin effect assumes that the constitutive relation between the current and the electric field is local. The anomalous skin effect appears when this assumption is not valid and the local current depends nonlocally on the field distribution. Note that the name anomalous skin effect should not be confused with the influence of chiral anomalies. Both the normal and anomalous skin effects appear when the interactions between electrons are weak. Third, the viscous skin effect manifests itself in clean materials when the interactions between charge carriers become large and the charge flow is described by hydrodynamics. It is an example of boundary layer phenomena in fluid mechanics that arise in the immediate vicinity of a bounding surface where the effects of viscosity are significant. 

The goal of this paper is to study skin effects for a Weyl metal, focusing on imprints of chiral anomalies.  Chiral anomaly leads to the appearance of the chiral magnetic current \cite{li_chiral_2016}
\be
\vct J_{\mathrm{CME}} \propto \tau_5 (\vct E\cdot \vct B)\vct B,
\label{cme}
\ee
where $\tau_5$ is the time needed for the relaxation of the axial charge (i.e., the difference between the charge densities of the particles with opposite chirality). In the standard treatment, which focuses on the dc conductivity, $\tau_5$ is related to internode scattering induced by impurities \cite{SonSpivak2013, Spivak2016, Burkov2014}. However, when an ac field is applied, different relaxation mechanisms may become dominant. 

In this work we provide a detailed study of the change of the chiral magnetic conductivity across a wide frequency spectrum. To this end, we generalize the chiral kinetic theory framework to account for different relaxation processes present in realistic materials. These processes include both momentum relaxing collisions coming from the scattering of quasiparticles with impurities as well as momentum conserving collisions between quasiparticles. We follow the path of expanding the collision kernel in the eigenfunctions of the angular momentum. Such a procedure was successfully applied in the context of two-dimensional materials \cite{Ledwith2019, Kiselev2020, Guo2017, Lucas2017, Lucas2018b}.

In the process we uncover a new transport regime, the existence of which depends on the presence of the chiral anomaly. In this regime the relaxation of momentum is dominated by scattering off impurities, but the relaxation of the axial charge in the skin layer takes place mainly through the diffusion of particles out of the layer and into the bulk of the material. This process causes a significant increase of the skin depth, as well as a different scaling of the surface impedance with frequency, in the presence of magnetic field. This observation gives a clear experimental imprint of the chiral anomalies in optical conductivity measurements.

This paper is organised as follows. In Sec. \ref{sec_transport}, we provide an overview of different transport regimes in metals and their relation to the skin effect, and we explain the physical origin of the newly discovered \textit{anomaly-induced nonlocal} regime. In Sec. \ref{sec_kinetic}, we introduce the equations of the semiclassical chiral kinetic theory with second-order corrections, which are commonly overlooked in the context of transport in topological semimetals. In Sec. \ref{sec_linearize}, we linearize the Boltzmann equation and construct a collision operator that includes the relevant relaxation mechanisms. In Sec. \ref{sec_conduct}, we solve the Boltzmann equation and find the local and nonlocal conductivities across a wide range of parameters. Finally, in Sec. \ref{sec_skin} we implement boundary conditions in order to calculate the skin depth and surface impedance for the different regimes, with a particular focus on the newly discovered regime.


\section{Transport regimes and the skin effect}
\label{sec_transport}

\begin{figure} 	            
\includegraphics[width=0.9\columnwidth]{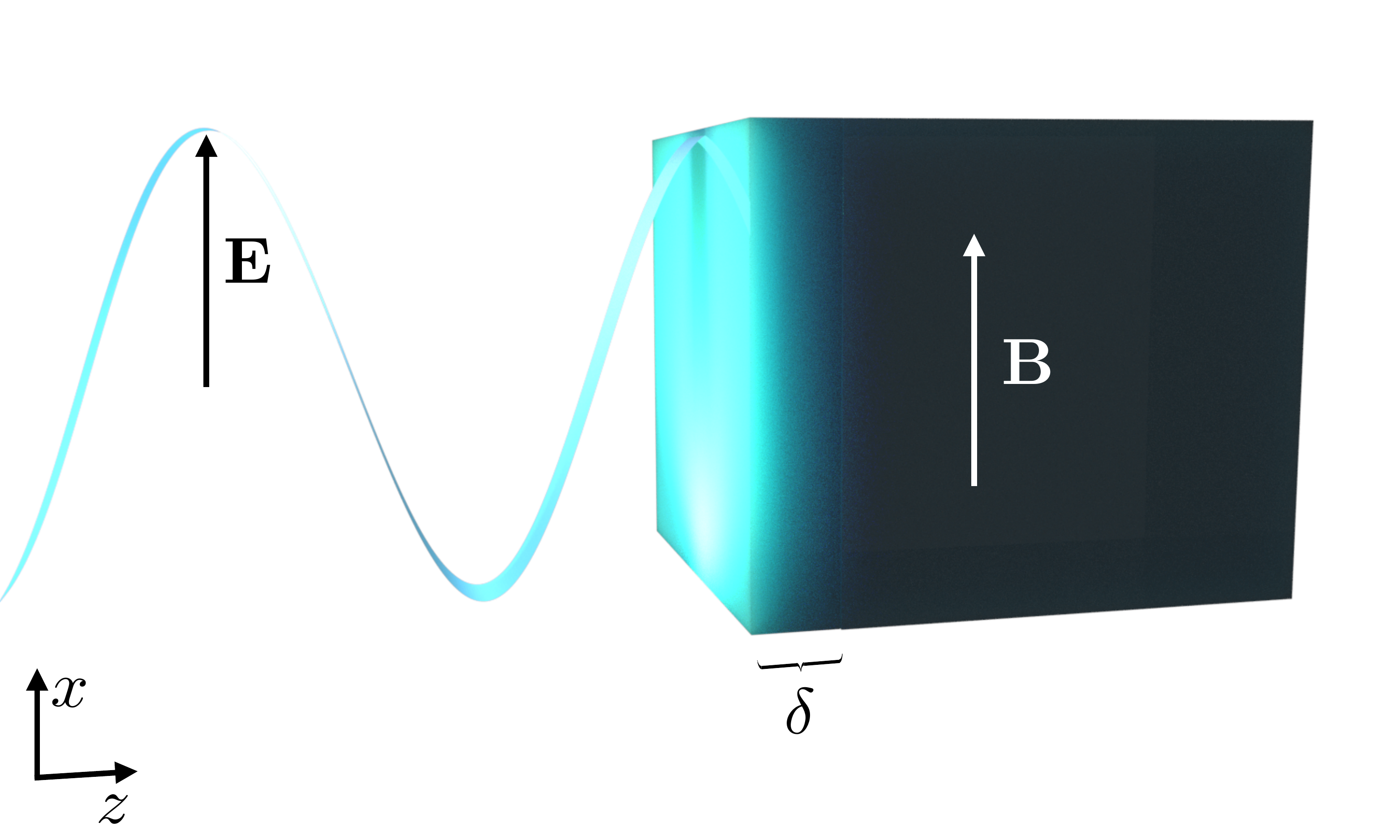}       
\caption{A schematic illustration of the skin effect. The propagation of an electromagnetic wave entering a conductor is limited to a layer of width $\delta$, called the skin depth. The skin depth changes with the frequency of the electric field $\omega$, making it possible to probe different transport regimes in a single sample of a Weyl semimetal. To determine the effect of the chiral anomaly on transport, the sample is placed in an external magnetic field $\vct B$ aligned parallel to the propagating electric field.}
\label{fig_setup}   
\end{figure} 

The skin effect is concerned with the flow of charge in a conductor under the influence of an oscillating electric field. It was first discovered in the context of spherical conductors \cite{lamb_electrical_1883} as a consequence of Maxwell's equations, which impose the propagation of electric field in a medium to be described by the equation
\be
\nabla^2 \vct E - \nabla\left( \nabla \cdot \vct E \right) - c^{-2}\partial_t^2 \vct E= \mu \partial_t \vct J,
\label{maxwell}
\ee
where $\vct J$ is the current density, $c$ is the speed of light, and $\mu$ is the magnetic permeability in the medium. When propagating inside a conductor, the electric field is attenuated by the induced currents and the magnitude of the field drops exponentially with the distance. This decay is characterized by a length $\delta$ called the skin depth (Fig. \ref{fig_setup}). To study the skin effect, one needs to supplement Maxwell's equations with a model describing the dynamics of particles in an electric field. The simplest model providing such dynamics is of the constitutive type and assumes a local relation between the current $\vct J(x, t)$ and the electric field $\vct E(x, t)$. This assumption is not always justified and nonlocal effects have to be taken into account, which can be done for almost free electrons by resorting to kinetic theory \cite{reuter_theory_1948}. Finally, in clean materials the relation between the current and the electric field can be hydrodynamic and controlled by the fluid viscosity. Kinetic theory can also be used to capture the effects of viscosity, providing a unified interpretation of these three regimes in terms of different scattering mechanisms in the conductor, which affects the final form of the optical conductivity \cite{Valentinis2021}. 

Electrons moving in the bulk of a (semi)metal experience different types of scattering: 1) electron-electron collisions, which conserve the total momentum, and 2) electron-impurity and electron-phonon collisions, which dissipate momentum. Correspondingly, we denote the mean free path between momentum-conserving collisions as $l_{\mathrm{mc}}$ and the mean free path for momentum-relaxing collisions as $l_{\mathrm{mr}}$. Other length scales essential to the problem are the skin depth $\delta$, as well as the path traversed by an electron over one period of oscillation of the field, $l_\omega \approx v_\mathrm{F}/\omega$, where $v_\mathrm{F}$ is the Fermi velocity and $\omega$ is the frequency of the driving electric field. Depending on the relative values of these length scales, one can distinguish the following different transport regimes:

(1) When $l_{\mathrm{mr}} \ll \delta$ or $l_\omega \ll \delta$, the electric field can be assumed to be uniform on the length of the free path of an electron, and consequently the relation between the electric field and electric current is local in space: $\vct J(\vct x, \omega) = \sigma(\omega)\vct E (\vct x, \omega)$. This is the regime where the usual Drude formula for optical conductivity can be used. The skin effect in this regime is called the \textit{normal skin effect} \cite{lamb_electrical_1883}. In this article we further differentiate between the \textit{low-frequency normal} regime (when $l_{\mathrm{mr}}$ is the shortest length scale) and \textit{high-frequency normal} regime (when $l_\omega$ is the shortest length scale).

(2) In the case of very pure samples at low temperatures, the electron-impurity and electron-phonon collisions can become less frequent than the electron-electron collisions; that is, $l_{\mathrm{mc}} \ll l_{\mathrm{mr}}$. Then, in some frequency range it may happen that $l_{\mathrm{mc}} \ll \delta \ll l_{\mathrm{mr}}$. In that case, the motion of the electron can be seen as a random walk with step size equal to $l_{\mathrm{mc}}$. The length of the path that an electron has to traverse to cross the skin layer is of the order $\delta^2/l_{\mathrm{mc}}$, so momentum relaxation in the skin layer happens on the length scale of $\mathrm{min}\{\delta^2/l_{\mathrm{mc}}, l_{\mathrm{mr}}\}$. If the condition $\delta^2/l_{\mathrm{mc}} \ll l_{\mathrm{mr}}$ is fulfilled, we are in the hydrodynamic regime, where the skin effect is known as the \textit{viscous skin effect}, first studied by Gurzhi \cite{Gurzhi1964}.

(3) Finally, when in some frequency range $\delta$ is the smallest length scale, i.e., $\delta \ll \mathrm{min}\{l_{\mathrm{mr}},l_{\mathrm{mc}},l_\omega\}$, we enter a regime in which electrons can be assumed not to undergo any scattering in the skin layer, so the motion of the electrons is ballistic. The skin effect in this regime is called the \textit{anomalous skin effect} \cite{reuter_theory_1948}.

In Weyl metals the presence of the chiral anomaly complicates this picture. When the material is placed in parallel electric and magnetic fields, the axial anomaly causes an imbalance between the densities of quasiparticles with different chiralities, and the chiral magnetic effect gives rise to the current in Eq. (\ref{cme}). Therefore, in addition to the rate of relaxation of momentum, also the rate of relaxation of the axial charge becomes relevant for the transport of electric charge. 

While at low frequencies both momentum and axial charge relaxation originate in the same physical processes, namely collisions with impurities, the relevant length scale for the latter, which we call $l_{\mathrm{inter}}$, can be much longer than $l_{\mathrm{mr}}$ \cite{SonSpivak2013, Spivak2016, Burkov2014}. Consequently, in a certain frequency range $l_{\mathrm{mr}}\ll\delta\ll l_{\mathrm{inter}}$ can occur. In analogy to our discussion of the hydrodynamic regime, in addition to the internode scattering, relaxation of the axial charge takes place also through diffusion. Since $l_{\mathrm{mr}}\ll \delta$, we are in the diffusive regime and the time needed for the charge imbalance to diffuse through the skin layer is of the order $\delta^2/\left(v_\mathrm{F} l_{\mathrm{mr}}\right)$. If the condition $\delta^2/l_{\mathrm{mr}} \ll l_{\mathrm{inter}}$ is satisfied, diffusion becomes the primary mechanism of the axial charge relaxation in place of the internode scattering. The characteristic feature of this new regime is that while the classical contribution to the conductivity can still be described by the local Drude formula, the contribution resulting from the quantum anomaly is nonlocal. Therefore, we call this regime the \textit{anomaly-induced nonlocal} (AIN) regime. The interplay between the classical and quantum contributions has non-trivial consequences for the skin effect.

\section{Kinetic theory with second order corrections}
\label{sec_kinetic}

Our aim is to calculate currents in a Weyl semimetal by solving the Boltzmann kinetic equation. Hereafter, we consider a Weyl semimetal composed of two Weyl nodes with opposite chirality and assume that there are no other bands contributing to the conductivity. Consequently, we consider two distribution functions $f^{(s)}$ labeled by the chirality of the node $s =\pm 1$. In the vicinity of the node with chirality s, and in the absence of external fields, we take the system to be described by the idealized low-energy Hamiltonian,
\be
H(\vct p) = sv\bm{\sigma}\cdot\vct p.
\label{hamiltonian}
\ee
For such a Hamiltonian, the dispersion relation for the conduction bands takes the simple form 
\be
\epsilon_0(\vct p) = v|\vct p|.
\label{dispersion_0}
\ee
Finally, our semiclassical approach assumes a system whose Fermi energy $\epsilon_F$ satitsies the inequalities $kT \ll \epsilon_\mathrm{F}$ and $\hbar\omega \ll \epsilon_\mathrm{F}$, such that there are no antiparticle excitations. 

The semiclassical equations of motion in the chiral kinetic theory are
\be
D^{(s)}\dot{\vct x}^{(s)} = \left[\vct v^{(s)}_\mathrm{M}  + e\hbar\vct E\times \vct \Omega^{(s)}
  + e \hbar\left(\vct v^{(s)}_\mathrm{M}\cdot \vct \Omega^{(s)}  \right)  \vct B    \right],
\label{dotx}
\ee
\be
D^{(s)}\dot{\vct p}^{(s)} =\left[e \vct E + e\vct v^{(s)}_\mathrm{M} \times\vct B + e^2\hbar (\vct E\cdot \vct B)\vct \Omega^{(s)} \right],
\label{dotp}
\ee
where $\vct v^{(s)}_{\mathrm{M}} = \partial_{\vct p} \epsilon^{(s)}_\mathrm{M}$ is the group velocity and
\be
D^{(s)} = 1+e\hbar \vct B\cdot \vct \Omega^{(s)}
\ee
is the volume element of the phase space in the presence of the Berry curvature $\vct \Omega^{(s)}$ \cite{Duval2006,Stephanov2012}. Charge and current densities are defined as
\be
\rho(\vct x, t) = e \sum_{s = \pm 1}\int \frac{d^3 p}{(2\pi\hbar)^3} D^{(s)}f^{(s)}(\vct x, \vct p, t),
\ee
\be
\vct J(\vct x, t) = e \sum_{s = \pm 1}\int \frac{d^3 p}{(2\pi\hbar)^3}D^{(s)}\dot{\vct x}^{(s)} f^{(s)}(\vct x, \vct p, t).
\label{J_def}
\ee
The distribution functions $f^{(s)}(\vct x, \vct p, t)$ can be determined from the corresponding Boltzmann equations
\be
\partial_t f^{(s)} + \dot{\vct x}^{(s)}\cdot\partial_{\vct x}f^{(s)}+\dot{\vct p}^{(s)}\cdot \partial_{\vct p}f^{(s)} = C^{(s)}[f^{(s)}, f^{(-s)}],
\label{boltzmann_general}
\ee
where $C^{(s)}$ is the collision integral (discussed in the next section). 

The presence of external fields changes the simple dispersion relation given in Eq. (\ref{dispersion_0}). In particular, corrections linear in $\vct B$, related to the orbital magnetic moment (OMM) of the wavepackets, have been shown to lead to a qualitative change in the magnetoconductance \cite{Knoll2020, Xiao2020}. As the chiral magnetic effect is seen on the order $\vct B^2$, corrections to the semiclassical equations of motion quadratic in $\vct B$ can also potentially affect the result. For that reason, in this work we include second-order corrections to the energy (see ref. \cite{Niu2015} and Appendix \ref{app_2nd})
\be
\epsilon_\mathrm{M}^{(s)}(\vct p) = v\left[|\vct p| - s\frac{e \hbar \vct B \cdot \vct p}{2 |\vct p|^2}+\frac{e^2\hbar^2|\vct B|^2}{8 |\vct p|^3}-\frac{e^2 \hbar^2\left(\vct B\cdot\vct p\right)^2}{8 |\vct p|^5}\right].
\label{energy}
\ee
The magnetic field also modifies the Berry curvature (see refs. \cite{Niu2014, Niu2015} and Appendix \ref{app_2nd}), which up to linear order in $\hbar$ reads
\be
\vct \Omega^{(s)}(\vct p) = s\frac{\vct{p}}{2 |\vct p|^3} - \frac{e \hbar \vct B}{4 |\vct p|^4} + \frac{e \hbar (\vct B\cdot \vct p)\vct p}{2 |\vct p|^6}.
\label{berry}
\ee
Using these formulas, the equations of motion are valid up to order $\vct B^2$. Second-order corrections to the equations of motion in the context of chiral kinetic theory were considered before in refs. \cite{Gorbar2017, Abbasi2019, Yang2020}.

It is important to state that the semiclassical approach is valid only for sufficiently weak magnetic fields. To be more precise, it is assumed that the dimensionless parameter 
\be
\alpha = \frac{e \hbar |\vct B| v^2}{2 \epsilon_\mathrm{F}^2},
\ee
satisfies $\alpha \ll 1$; otherwise the analysis would have to take into account the quantization of the Landau levels. Thus, $\alpha$ quantifies the magnitude of quantum corrections to the classical equations of motion.

\section{Linearized Boltzmann equation}
\label{sec_linearize}

The collision term $C^{(s)}[f^{(s)}, f^{(-s)}]$ in Eq. (\ref{boltzmann_general}) is zero when both particle species follow the Fermi-Dirac distribution; i.e., $C^{(s)}[f_0^{(s)}, f_0^{(-s)}]=0$ with
\be
f_0^{(s)}(\vct p) = \left[\exp\left(\frac{\epsilon_\mathrm{M}^{(s)}(\vct p) - \epsilon_\mathrm{F}}{kT}\right) +1\right]^{-1}.
\ee
To solve the Boltzmann equation, $f^{(s)}$ is expanded around $f_0^{(s)}$. Setting $f^{(s)}=f_0^{(s)}+f_1^{(s)}$ in Eq. (\ref{boltzmann_general}) shows that the leading-order solution $f_0$ contributes a term proportional to $\partial_{\vct p}f_0^{(s)} = \partial_{\vct p}\epsilon^{(s)}_\mathrm{M} \cdot \partial_{\epsilon_\mathrm{M}} f_0^{(s)}$ to the equation. Consequently, we parametrize the full solution as follows:
\be
f^{(s)} =  f_0^{(s)}(\vct p) + \partial_{\epsilon_\mathrm{M}} f_0^{(s)} \cdot \eta^{(s)}(\vct x, \vct p, t).
\label{etaexpansion}
\ee
For $kT \ll \epsilon_\mathrm{F}$, $\partial_{\epsilon_\mathrm{M}} f_0^{(s)} \approx - \delta\left[\epsilon_\mathrm{M}^{(s)}(\vct p) - \epsilon_\mathrm{F}\right]$, which means that we only have to pay attention to $\eta^{(s)}(\vct x, \vct p, t)$ with momenta $\vct p$ close to the Fermi surface. The shape of the Fermi surface is determined by equating the RHS of Eq. (\ref{energy}) to $\epsilon_\mathrm{F}$ and solving for $|\vct p|$. To that end, we introduce the unit vectors pointing in the direction of magnetic field and momentum,
\be
\hat{\vct b} = \vct B/|\vct B|,~~~~~~~~~~\hat{\vct p} = \vct p/|\vct p|,
\ee
respectively. The Fermi momentum in the direction $\hat{\vct p}$ can then be written, up to the second order in $\vct B$, as:
\be
p_\mathrm{F}^{(s)}(\hat{\vct p}) = \frac{\epsilon_\mathrm{F}}{v}\left[1+s\alpha(\hat{\vct p}\cdot\hat{\vct b})-\frac{1}{2}\alpha^2 - \frac{1}{2}\alpha^2(\hat{\vct p}\cdot\hat{\vct b})^2\right].
\label{fermi_mom}
\ee
Using this result it is possible to evaluate the Berry curvature $\vct \Omega^{(s)}$ and the group velocity $\vct v_\mathrm{M}^{(s)}$ at the Fermi surface up to the second order in $\alpha$:
\be
\vct\Omega^{(s)}(p_\mathrm{F}^{(s)}(\hat{\vct p}), \hat{\vct p}) = \frac{1}{e\hbar|\vct B|}\left[s \alpha \hat{\vct p} - \alpha^2 \hat{\vct b}\right],
\label{omega_s}
\ee
\be
\begin{split}
\vct v_\mathrm{M}^{(s)}(p_\mathrm{F}^{(s)}(\hat{\vct p}), \vct{\hat p})  = v & \left[\vct{\hat p} - s\alpha \vct{\hat b}+2 s \alpha (\hat{\vct p}\cdot\hat{\vct b})\hat{\vct p}-\frac{3}{2}\alpha^2\vct{\hat p}\right. \\
& \left.+\alpha^2 (\hat{\vct p}\cdot\hat{\vct b})\hat{\vct b}-\frac{3}{2}\alpha^2(\hat{\vct p}\cdot\hat{\vct b})^2\hat{\vct p}\right].
\end{split}
\label{vM_s}
\ee
In what follows we make the dependence on $p_\mathrm{F}$ implicit; i.e., $\vct \Omega^{(s)}(p_\mathrm{F}^{(s)}(\vct{\hat{p}}),\vct{\hat{p}}) \equiv \vct \Omega^{(s)}(\vct{\hat{p}})$, $\vct v_\mathrm{M}^{(s)}(p_\mathrm{F}^{(s)}(\vct{\hat{p}}),\vct{\hat{p}}) \equiv \vct v_\mathrm{M}^{(s)}(\vct{\hat{p}})$, $\eta^{(s)}(p_\mathrm{F}^{(s)}(\vct{\hat{p}}), \vct{\hat{p}}) \equiv \eta^{(s)}(\vct{\hat{p}})$.

Next, we turn our attention to the collision term on the RHS of Eq. (\ref{boltzmann_general}). We linearize this term by introducing the collision operator $\hat{C}^{(s)}$ defined as
\be
\hat{C}^{(s)}[\eta^{(s)},\eta^{(-s)}] \equiv \left( \partial_{\epsilon_\mathrm{M}} f_0^{(s)} \right)^{-1} C^{(s)}[f^{(s)},f^{(-s)}],
\label{collision_general}
\ee
where $f^{(s)} = f_0^{(s)} + \partial_{\epsilon_\mathrm{M}} f_0^{(s)} \cdot \eta^{(s)}$. At first, we neglect possible interactions between the nodes and consider properties of the collision operator acting on a well-defined chiral state $\hat{C}^{(s)}[\eta^{(s)}] \equiv \hat{C}^{(s)}[\eta^{(s)},0]$. This collision operator can be shown to be hermitian and nonpositive with respect to the inner product \cite{Ferziger, Cercignani, Ledwith2019}
\be
\langle \eta |\zeta \rangle^{(s)} = -\int \frac{d^3 p}{(2\pi\hbar)^3} D^{(s)}(\vct p)\left(\partial_{\epsilon_\mathrm{M}} f_0^{(s)}\right) \eta(\vct p)^* \zeta(\vct p).
\label{inner}
\ee
In the absence of an external magnetic field, $\vct B = 0$, the Fermi surface is a sphere and $D^{(s)}(\vct p) = 1$. Then, if we assume that the scattering in the bulk is isotropic, the problem is spherically symmetric and the eigenfunctions of $\hat{C}^{(s)}$ are the spherical harmonics
\be
\hat{C}^{(s)}[Y^m_l(\vct{\hat p})] = - \Gamma_{l,m}Y^m_l(\vct{\hat p})~~~~(\vct B = 0).
\label{collision0}
\ee
The eigenvalues $\Gamma_{l,m}$ describe the relaxation of the different modes. For example, if the distribution function $\eta^{(s)}$ is expanded in spherical harmonics as 
\be
\eta^{(s)} = \sum_{l,m}X_l^{m(s)}Y^m_l(\vct{\hat p}),
\ee
where $X_l^{m(s)} = X_l^{m(s)}(\vct x, t)$ are some functions of position and time, the rate of change of the charge density is
\be
\begin{split}
\frac{d \rho^{(s)}}{d t } &= e\int \frac{d^3 p}{(2 \pi \hbar)^3} \frac{d f^{(s)}(\vct x, \vct p, t)}{d t}  \\
& = e\int \frac{d^3 p}{(2 \pi \hbar)^3} \partial_{\epsilon_\mathrm{M}}f_0^{(s)}\cdot \hat{C}^{(s)}[\eta^{(s)}] \\
& = -\frac{e\epsilon_\mathrm{F}^2}{(2\pi)^2 \sqrt{\pi}\hbar^3 v^3 }\Gamma_{0,0} \left(-X_0^{0(s)}\right).
\end{split}
\ee
A simple calculation shows that $\rho^{(s)}\propto - X_0^{0(s)}$. Thus, $\Gamma_{0,0}$ describes the rate of the particle number relaxation. Similarly, using the fact that $\hat{p}_z \propto Y_1^0(\hat{\vct p})$, $\hat{p}_x \propto \left(Y_1^{-1}(\hat{\vct p})-Y_1^{1}(\hat{\vct p})\right)$, and $\hat{p}_y \propto i\left( Y_1^{-1}(\hat{\vct p})+Y_1^{1}(\hat{\vct p})\right)$, it is easy to see that $\Gamma_{1,M}$, where $M=-1,0,1$, describe the rates of relaxation of momentum. 

In the case of nonzero magnetic field, we can no longer use the spherical symmetry to find the eigenfunctions of the collision operator. Nevertheless, we can assume that they should be of the form $K_l^{m(s)} = Y_l^m + O(\alpha)$ and that the corresponding eigenvalues, $\Gamma_{0,0}$ and $\Gamma_{1,M}$, should again describe the rates of change of $\rho$ and $\langle \vct p\rangle$ respectively. The details of finding the functions $K_l^{m(s)}$ can be found in Appendix \ref{app_eigenfunctions}. In this work we expand $K_l^{m(s)}$ in spherical harmonics to the second order in $\alpha$.

At last, let us construct a collision operator that includes the relevant relaxation processes. One of them is the internode scattering, which transfers particles between the two nodes, leading to a dissipation of the axial charge; we denote the frequency at which this process happens by $\Gamma_\mathrm{inter}$. Another process is the intranode scattering that, together with the internode scattering, relaxes momentum; we denote the total rate at which momentum is dissipated by $\Gamma_\mathrm{mr}$ and, consequently, we take $\Gamma_{1,M} = \Gamma_\mathrm{mr}$ for $M=-1,0,+1$. Finally, we include in our analysis electron-electron scattering, which conserves momentum. Because modes with a high angular momentum do not contribute to the current, and in order to simplify the calculations, we assume a constant relaxation rate $\Gamma_\mathrm{tot}$ for all modes with the total angular momentum $L \geq 2$. All the scattering mechanisms that we consider, i.e., the inter- and intranode scattering, as well as the electron-electron collisions, contribute to $\Gamma_\mathrm{tot}$. On the basis of the physical origin of the relaxation rates we take $\Gamma_\mathrm{inter}<\Gamma_\mathrm{mr}<\Gamma_\mathrm{tot}$. Introducing the operators
\be
\begin{split}
P_0^{(s)}  & = |K_0^{0(s)}\rangle\langle K_0^{0(s)}|-|K_0^{0(s)}\rangle\langle K_0^{0(-s)}|, \\
P_1^{(s)} & = \sum_{M=-1,0,1}|K_1^{M(s)}\rangle\langle K_1^{M(s)}|,\\
P_{\mathrm{higher}}^{(s)}  & = 1 - |K_0^{0(s)}\rangle\langle K_0^{0(s)}| - P_1^{(s)},
\end{split}
\label{projectors}
\ee
where $s$ defines the distribution function which the operator acts on, the collision operator can be written as
\be
\hat{C}^{(s)} =  -\Gamma_\mathrm{inter} P_0^{(s)}  - \Gamma_\mathrm{mr} P_1^{(s)} - \Gamma_\mathrm{tot} P_{\mathrm{higher}}^{(s)}.
\label{collision}
\ee
Similar expansions of the collision operator were used in refs. \cite{Guo2017, Lucas2017, Lucas2018b, Kiselev2020} for different systems. It can be verified that with this choice of collision operator $\frac{d}{dt}\rho^{(s)} \propto -\Gamma_\mathrm{inter}\left(\rho^{(s)}- \rho^{(-s)}\right)$ and, in the absence of external force and internode scattering, $\frac{d}{dt}\langle \vct p\rangle^{(s)} \propto - \Gamma_\mathrm{mr} \langle \vct p\rangle^{(s)}$.

We are now in a position to write down the linearized Boltzmann equation [Eq. (\ref{boltzmann_general})]. Recalling that we need to compute $\eta^{(s)}$ up to linear order in $\vct E$ and quadratic in $\vct B$, we can neglect the anomalous Hall term in the spatial-derivative term
\begin{multline}
D^{(s)}(\vct{\hat p})\dot{\vct x}^{(s)}\cdot\partial_{\vct x}f^{(s)} \\
=\partial_{\epsilon_\mathrm{M}} f_0^{(s)}\left[\vct v_\mathrm{M}^{(s)} + e \hbar\left(\vct v_\mathrm{M}^{(s)}\cdot \vct \Omega^{(s)}  \right)  \vct B    \right]\cdot\partial_{\vct x}\eta^{(s)},
\end{multline}
while the relevant contributions of the momentum-derivative term are
\be
\begin{split}
D^{(s)}(\vct{\hat p})&\dot{\vct p}^{(s)}\cdot\partial_{\vct p}f^{(s)} \\
& =\left[e \vct E + e^2\hbar (\vct E\cdot \vct B)\vct \Omega^{(s)}     \right]\cdot \vct v_\mathrm{M}^{(s)} \partial_{\epsilon_\mathrm{M}} f_0 \\
& + e \left[\vct v_\mathrm{M}^{(s)} \times\vct B\right]\cdot\partial_{\vct p}\eta^{(s)} \partial_{\epsilon_\mathrm{M}} f_0^{(s)}.
\end{split}
\label{momentum_term}
\ee
Note that we use the formulas for $\vct \Omega^{(s)}$ and $\vct v_\mathrm{M}^{(s)}$ given in Eqs. (\ref{omega_s}) and (\ref{vM_s}), respectively, in order to account for the second-order corrections to the equations of motion. The last term on the RHS of Eq. (\ref{momentum_term}) results from the Lorentz force and is the source of the classical Hall effect and classical magnetoresistance. Because these phenomena are well known and this article focuses on quantum corrections to the longitudinal conductivity, we do not include them here. Additionally, we are going to consider a setup where the electric field is parallel to the (external) magnetic field, in which case the classical magnetoresistance vanishes in the normal regime. On the other hand, in the hydrodynamic and ballistic regimes the situation is more complicated: for details, see Appendix \ref{app_estimate}. These considerations allow us to write the linearized Boltzmann equation
\be
\begin{split}
 D^{(s)}\partial_t\eta^{(s)}& + \left[\vct v^{(s)}_\mathrm{M} + e \hbar\left(\vct v^{(s)}_\mathrm{M}\cdot \vct \Omega^{(s)}  \right)  \vct B    \right]\cdot\partial_{\vct x}\eta^{(s)} \\
& +e \left[\vct E\cdot \vct v^{(s)}_\mathrm{M} + e\hbar (\vct E\cdot \vct B)\left(\vct \Omega^{(s)}\cdot \vct v^{(s)}_\mathrm{M}\right)     \right] \\
& =D^{(s)}\hat{C}^{(s)}[\eta^{(s)},\eta^{(-s)}],
\end{split}
\label{boltzmann_expanded}
\ee
with $\hat{C}^{(s)}$ defined in Eq. (\ref{collision}).

\section{Bulk conductivities}
\label{sec_conduct}

In this section, we calculate the current flowing through the bulk of a Weyl semimetal subject to a weak ac electric field. The material is modelled as an infinite half-space $z>0$, while $z<0$ is the vacuum. We place the material in a static external magnetic field parallel to the surface, $\vct B = B\hat{\vct x}$, and consider an oscillating electric field propagating through the bulk and polarized in the same direction as the magnetic field: $\vct E(z, t) = E(z,t)\vct{\hat x}$. The electric field is assumed to be uniform in the $xy$ plane.

The general expression for the current, after taking into account the orbital magnetization, can be written as
\be
\vct J = \vct J_{\mathrm{kin}} + \vct J_{\mathrm{magn}},
\ee
where $\vct J_{\mathrm{kin}}$ is given by Eq. (\ref{J_def}) and represents the ``kinematic" part resulting from the forward motion of the wavepackets, while
\be
\vct J_{\mathrm{magn}} = \nabla \times \sum_{s=\pm 1} \int \frac{d^3 p}{(2\pi)^3}D^{(s)}(\vct p)\frac{s e \hbar v}{2 |\vct p|} \hat{\vct p} f^{(s)}(q,\vct p)
\label{J_magn}
\ee
is the magnetization current. However, in the setup that we consider $\vct J_{\mathrm{magn}}=0$ (see Appendix \ref{app_estimate} for a proof). Consequently, in what follows
\be
\vct J = \vct J_{\mathrm{kin}}.
\ee
Equation (\ref{J_def}) is now expanded to the second order in $\alpha$. In this article we neglect the anomalous Hall effect, i.e., the $e\hbar\vct E\times \vct \Omega^{(s)}$ term. In a time-reversal symmetry-breaking Weyl semimetal with a single pair of nodes this can be justified, if the electric field is aligned parallel (or near-parallel) to the vector connecting the Weyl nodes in reciprocal space. After switching to spherical coordinates $(p, \theta, \phi)$ in momentum space, one obtains
\be
\begin{split}
\vct J(z) & = e\sum_{s = \pm 1}\int \frac{d^3 p}{(2\pi\hbar)^3}D(\vct p)\dot{\vct x}(\vct p)(-\delta\left[\epsilon_\mathrm{M}(\vct p) - \epsilon_\mathrm{F}\right])\eta(z, \vct p) \\
& = -\frac{e \epsilon_\mathrm{F}^2}{v^2(2\pi\hbar)^3}\sum_{s = \pm 1}\int d(\cos\theta)d\phi \\
& \left[\vct{\hat p} + 3s\alpha (\hat{\vct p}\cdot\hat{\vct b})\vct{\hat p} -\alpha^2\vct{\hat p} + \alpha^2 (\hat{\vct p}\cdot\hat{\vct b})\vct{\hat b}\right]\eta^{(s)}(z, \vct{\hat p}),
\end{split}
\label{J_expanded}
\ee
where we suppress the $(s)$ labels on all quantities in the first line.

To find the current from Eq. (\ref{J_expanded}), the distribution functions $\eta^{(s)}(z, \vct p)$ have to be determined from the Boltzmann equation [Eq. (\ref{boltzmann_expanded})]. We shall work with Fourier-transformed quantities defined as follows:
\be
\begin{split}
\eta^{(s)}(z,\vct p, t) & = \int d\omega dq e^{i\omega t - i q z} \eta^{(s)}(q,\vct p,\omega), \\
\vct E(z,\vct p, t) & = \int d\omega dk e^{i\omega t - i q z} \vct E(q,\vct p,\omega).
\end{split}
\label{fourier}
\ee
The Boltzmann equation is solved perturbatively in $\alpha$ after expanding the out-of-equilibrium distribution in spherical harmonics as
\be
\eta^{(s)} = \sum_{l,m}\left(A_l^m + s\alpha B_l^m + \alpha^2 C_l^m\right)Y_l^m(\vct{\hat p}).
\label{eta_expanded}
\ee
Next, the collision integral on the RHS of Eq. (\ref{boltzmann_expanded}) is evaluated using the definition of the inner product (\ref{inner}), the expansion of the collision operator into projection operators (\ref{projectors}, \ref{collision}), and the form of the eigenvectors $K_l^{m(s)}$ determined in Appendix \ref{app_eigenfunctions}. Then, Eq. (\ref{boltzmann_expanded}) is projected onto a spherical harmonic of degree L and order M by multiplying both sides by $Y_L^{M*}(\hat{\vct p})$ and integrating over the angles. These integrals can be evaluated analytically, producing an infinite system of equations labeled by $L, M$. Finally, terms that are of the same order in $\alpha$ are equated, leading to three systems of equations (for terms of order $1$, $\alpha$ and $\alpha^2$) that are solved in turn.

We shall now solve the system of equations corresponding to the lowest level of approximation ($\alpha = 0$) in order to demonstrate the general strategy of solution, which is in the same spirit as the one used in ref. \cite{Lucas2018c} for a 2D system. At this level, the equations are \footnote{The factor of $1/2$ in front of the bracket in the last equation is in fact an approximate value, but this approximation works very well here. For a discussion of this approximation, see Appendix \ref{app_solution}.}
\be
i\omega A_0^0 -i q v  \frac{\sqrt{3}}{3} A_1^0 = 0,
\label{aux00}
\ee
\be
\left(i\omega+\Gamma_\mathrm{mr}\right) A_1^0 -i q v \left[\frac{\sqrt{3}}{3} A_0^0+\sqrt{\frac{4}{15}} A_2^0\right]=0,
\label{aux10}
\ee
\be
(i\omega+\Gamma_\mathrm{mr}) A_1^{\pm1} -iq v \frac{\sqrt{5}}{5} A_2^{\pm1} \mp e v \sqrt{\frac{2\pi}{3}} E= 0,
\label{aux11}
\ee
\be
\left(i\omega+\Gamma_\mathrm{tot}\right) A_L^M -i q v \frac{1}{2} \left[A_{L-1}^M + A_{L+1}^M\right] = 0 ~~~~~~(\mathrm{for}~L \geq 2).
\label{aux2}
\ee
Note that in the above, due to the cylindrical symmetry of the system in the absence of an electric field, only equations for a fixed M are coupled. Equation (\ref{aux2}) is a recurrence relation that can be solved by the ansatz $A_L^M = r^{L-1}A_1^M$. There are two solutions, from which we choose the one that ensures convergence of the series,
\be
r = -\frac{i}{qv} \left[\Gamma_\mathrm{tot} + i \omega - \sqrt{\left(\Gamma_\mathrm{tot}+i\omega\right)^2 + q^2 v^2}\right],
\label{r}
\ee
where we take the square root in the bracket to return a value with a positive real and positive imaginary part. Inserting $A_2^{M} = rA_1^{M}$ for $M=-1,0,+1$ in Eqs. (\ref{aux00})--(\ref{aux11}) yields a closed system of equations that can easily be solved, while for $|M| > 1$ the solution is trivial: $A_L^M = 0$. This way, the full distribution function at the classical level is found to be
\be
\begin{split}
A_1^{\pm 1} &= \pm \sqrt{\frac{2\pi}{3}}\frac{e v E}{\Gamma_\mathrm{mr}+i\omega-\frac{\sqrt{5}}{5}iqvr},\\
A_L^{\pm 1} &= r^{L-1}A_1^{\pm 1},~~~~~~A_L^{M\neq \pm1}=0.
\end{split}
\label{classical_solution}
\ee

The same procedure can be used to find corrections to $\eta^{(s)}$ at higher orders in $\alpha$ as well. First, the solution for $A_L^M$ is plugged into the equations that are linear in $\alpha$ in order to find $B_L^M$. Subsequently, the equations for high L are expressed as a recurrence relation that is then solved, and its solution is used to turn the system of equations for $L=0,1,2$ into a closed system. After solving this system, the solutions for $A_L^M$ and $B_L^M$ are plugged into the quadratic-order equations to find $C_L^M$ and the procedure is repeated. As the calculations are lengthy and not particularly instructive, we relegate the details to Appendix \ref{app_solution}. 

\begin{figure} 	            
\includegraphics[width=1.0\columnwidth]{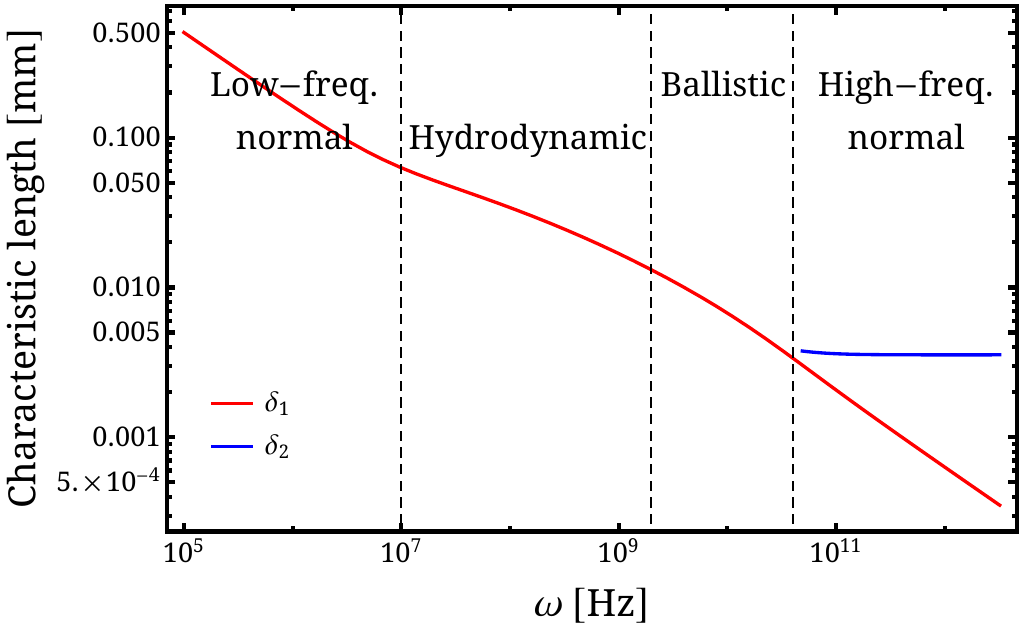}       
\caption{Dependence of the characteristic length scales at which a propagating electric field decays on $\omega$. The two curves represent two solutions of Maxwell's equations in the form of Eq. (\ref{classical_maxwell}) supplemented by the formula for the classical conductivity Eq. (\ref{classical_cond}) and experimentally realistic parameters (main text): $v=1.4\times10^5$ m/s, $\epsilon_\mathrm{F}=20$ meV, $\Gamma_\mathrm{mr}= 10^9$ Hz, $\Gamma_\mathrm{tot}= 10^{10}$ Hz. We select only solutions satisfying $\mathrm{Re}[q_i]>0$, $\mathrm{Im}[q_i]<0$ and plot their characteristic decay lengths, defined as $\delta_i = -\mathrm{Im}[q_i]^{-1}$. At low frequencies there exists only one solution $q_1$ satisfying these conditions. We can observe three transport regimes (identified as low-frequency normal, hydrodynamic and ballistic) where $\delta_1$ shows different scaling with $\omega$, and the cross-overs (marked with blue dashed lines) happen around the values of $\omega$ for which $|q_1 v|\approx\Gamma_\mathrm{mr}$ and $|q_1 v| \approx \Gamma_\mathrm{tot}$. Above a certain value of $\omega$, around the point when $|q_1 v| \approx \omega$, another solution $q_2$ becomes relevant. Since $\delta_2$ is the longer of the two decay lengths, it dominates the propagation of the electric field and we are in what we call the high-frequency normal regime.}
\label{regimes}   
\end{figure} 

After finding the distributions $\eta^{(s)}(q,\vct{\hat p}, \omega)$, the current is evaluated using Eq.~(\ref{J_expanded}). The current can be expressed as $\vct J(q,\omega)=\sigma(q,\omega)\vct E(q,\omega)$, where the conductivity $\sigma$ is a complicated function of $q$ and $\omega$. Here we only present the formulas for the conductivity in the limiting cases, in which some of the frequency scales involved in the problem are much greater than the others. The relevant frequency scales are the relaxation rates, which satisfy the inequalities $\Gamma_\mathrm{inter}<\Gamma_\mathrm{mr}<\Gamma_\mathrm{tot}$, $\omega$ (the driving frequency) and $q v$. As both $q v$ and $\omega$ can in general be placed between any two relaxation rates in this chain of inequalities, there is in principle a large number of limiting cases that can be considered. Instead of analysing each one of them, we focus only on the regimes that could be accessed in a skin effect experiment, which we identify in the following way. We choose certain realistic material parameters, close to those reported for WP$_2$ \cite{Kumar2017, Gooth2018}: $v=1.4\times10^5$ m/s, $\epsilon_\mathrm{F}=20$ meV, $\Gamma_\mathrm{mr}= 10^9$ Hz, $\Gamma_\mathrm{tot}= 10^{10}$ Hz. We moreover assume that $\Gamma_\mathrm{inter}$ is two orders of magnitude smaller that $\Gamma_\mathrm{mr}$ as reported for TaAs \cite{zhang_signatures_2016}, so $\Gamma_\mathrm{inter}=10^7$ Hz. Next, let us note that the classical conductivity can be calculated using Eq. (\ref{classical_solution}) and Eq. (\ref{J_expanded}) evaluated at the classical (i.e., $\alpha=0$) level, yielding
\be
\sigma_{\mathrm{cl}}(q,\omega) = 2\,\varepsilon\,\omega_\mathrm{P}^2 \frac{1}{\Gamma_\mathrm{mr} + i\omega - \frac{\sqrt{5}}{5}iqvr},
\label{classical_cond}
\ee 
where the plasma frequency $\omega_\mathrm{P}$ is defined by
\be
\omega_\mathrm{P}^2 = \frac{e^2 \epsilon_\mathrm{F}^2}{6 \pi^2 \varepsilon \hbar^3 v}
\ee
and $\varepsilon$ is the electric permittivity of the medium. Then, we can solve the Fourier-transformed Eq. (\ref{maxwell})
\be
\left(-q^2 +\frac{\omega^2}{c^2}\right) = i\,\mu\,\omega\,\sigma(q,\omega)_{\mathrm{cl}}
\label{classical_maxwell}
\ee
to obtain the wavectors $q$, which correspond to modes that can propagate through the bulk, as a function of $\omega$. This equation has in general six solutions, but we can restrict our attention to those with a negative imaginary and positive real part, which in accordance with Eq. (\ref{fourier}) correspond to modes propagating in the positive direction along $z$ and decaying with the distance. It turns out that out of the six solutions one satisfies this condition for all $\omega$, and one satisfies it at high frequencies only; we denote these solutions $q_1$ and $q_2$, respectively. The characteristic length scales at which these modes decay, defined as $\delta_i = -\mathrm{Im}[q_i]^{-1}$, can be used as a proxy for the classical skin depth and they are plotted in Fig. \ref{regimes}. Based on the different scaling of $q_i$ with $\omega$ we can identify four relevant regimes, which we call low-frequency normal, hydrodynamic, ballistic, and high-frequency normal. In all these regimes, except the last one, $\omega$ is smaller than all the relevant frequency scales and can thus be neglected in the analysis.

When quantum effects are included, the low-frequency normal regime splits into two, as discussed in Sec. \ref{sec_transport}. The different regimes and the corresponding conductivities are assembled in Table \ref{table_cond}. It can be seen that while in both the low-frequency normal and the anomaly-induced nonlocal regimes the relaxation of momentum and axial charge are related to different frequency scales, in the other regimes they both happen at a comparable rate.

\begin{table*}[]
\begin{tabular}{|l|c|c|c|}
\hline
Regime & Valid when & Conductivity $\sigma(q, \omega)$ & Surface impedance $Z(\omega)$\\
\hline 
\begin{tabular} {@{}l@{}} Low-frequency \\normal  \end{tabular}&\begin{tabular} {@{}c@{}} $q^2 v^2 \ll \Gamma_\mathrm{inter}\Gamma_\mathrm{mr}$\\$\omega\ll \Gamma_\mathrm{inter}$ \end{tabular} & $\varepsilon \,\omega_\mathrm{P}^2 \left[\frac{2}{\Gamma_\mathrm{mr}} + \alpha^2\frac{3}{\Gamma_\mathrm{inter}}-\alpha^2\frac{66}{15}\frac{1}{\Gamma_\mathrm{mr}}\right]$ & $e^{i\pi/4}\mu\sqrt{\frac{\omega c^2}{\omega_\mathrm{P}^2}}\left(\frac{2}{\Gamma_\mathrm{mr}} + \alpha^2\frac{3}{\Gamma_\mathrm{inter}}-\alpha^2\frac{66}{15}\frac{1}{\Gamma_\mathrm{mr}}\right)^{-\frac12}$ \\ 
\hline
\begin{tabular} {@{}l@{}} Anomaly-induced \\ nonlocal \end{tabular} & \begin{tabular} {@{}c@{}} $\Gamma_\mathrm{inter}\Gamma_\mathrm{mr}\ll q^2 v^2 \ll \Gamma_\mathrm{mr}^2$\\$\omega\ll q^2v^2/\Gamma_\mathrm{mr}$ \end{tabular} & $\varepsilon \,\omega_\mathrm{P}^2 \left[\frac{2}{\Gamma_\mathrm{mr}} + \alpha^2\frac{18\Gamma_\mathrm{mr}}{q^2v^2}\right]$ &  see Eq. (\ref{mixed_impedance}) in Sec. \ref{sec_skin}\\ 
\hline
Hydrodynamic & \begin{tabular} {@{}c@{}} $\Gamma_\mathrm{mr}\Gamma_\mathrm{tot}\ll q^2 v^2 \ll \Gamma_\mathrm{tot}^2$\\$\omega\ll q^2v^2/\Gamma_\mathrm{tot}$ \end{tabular} &  $\varepsilon\, \omega_\mathrm{P}^2\frac{\Gamma_\mathrm{tot}}{q^2 v^2}\left[9.0-26\alpha^2\right]$ & $\frac{1}{2}\left(e^{i\pi/8}+e^{i5\pi/8}\right)\mu\left(\frac{\omega^3v^2c^2}{\omega_\mathrm{P}^2\Gamma_\mathrm{tot}\left[9.0-26\alpha^2\right]}\right)^{1/4}$\\ 
\hline
Ballistic & \begin{tabular} {@{}c@{}} $\Gamma_\mathrm{tot}\ll q v$\\$\omega\ll qv$ \end{tabular} & $\varepsilon\, \omega_\mathrm{P}^2\frac{1}{q v}\left[4.5 +4.3\alpha^2\right]$ & $\frac{4\sqrt{3}}{9}e^{i\pi/3}\mu\left(\frac{\omega^2vc^2}{\omega_\mathrm{P}^2\left[4.5 +4.3\alpha^2\right]}\right)^{1/3}$\\
\hline
\begin{tabular} {@{}l@{}} High-frequency \\ normal \end{tabular} & \begin{tabular} {@{}c@{}} $qv \ll \omega$\\$\Gamma_\mathrm{tot}\ll\omega$ \end{tabular} & $-i\varepsilon\, \omega_\mathrm{P}^2\frac{1}{\omega}\left[2+\frac{8}{5}\alpha^2\right]$ & $-i\mu \frac{\omega c}{\omega_\mathrm{P}} \left(2+\frac{8}{5}\alpha^2\right)^{-\frac12}$\\
\hline
\end{tabular}
\caption{Conductivity and surface impedance in the different regimes and their regions of validity. The numerical coefficients in the hydrodynamic and ballistic regimes are rounded up to two significant figures.}
\label{table_cond}
\end{table*}

We can extract further information as to the origin of the numerical coefficients in Table \ref{table_cond} if before the start of the calculations we multiply the term corresponding to the OMM in Eq. (\ref{energy}) (that is, $- s\frac{e \hbar \vct B \cdot \vct p}{2 |\vct p|^2}$) by a factor $\xi_1$, and similarly we multiply all the terms resulting from the second-order corrections to the energy and Berry curvature in Eqs. (\ref{energy}, \ref{berry}) by $\xi_2$, and then we keep track of these coefficients during the computations. This allows us to determine the following:
\begin{itemize}
\item The second-order corrections to the equations of motion have no influence on the conductivity either in the normal or the AIN regimes, affecting only the hydrodynamic and ballistic results.
\item For a relatively high $\Gamma_\mathrm{inter}$, namely $\Gamma_\mathrm{inter}>15 \Gamma_\mathrm{mr}/22\approx0.7\Gamma_\mathrm{mr}$, the magnetoconductivity in the low-frequency regime changes sign from positive to negative, in qualitative agreement with refs. \cite{Knoll2020, Xiao2020}. This is found to be caused by the contribution of the OMM, as setting $\xi_1=0$ and neglecting electron-electron collisions (so that $\Gamma_\mathrm{tot}=\Gamma_\mathrm{mr}$) would result in a different formula for the conductivity, denoted here as $\tilde{\sigma}$,
\be
\tilde{\sigma}(q,\omega) = \varepsilon\, \omega_\mathrm{P}^2 \left[\frac{2}{\Gamma_\mathrm{mr}} + \alpha^2\frac{3}{\Gamma_\mathrm{inter}}-\alpha^2\frac{42}{15}\frac{1}{\Gamma_\mathrm{mr}}\right],
\ee
and the magnetoconductivity would always be positive, again in agreement with refs. \cite{Knoll2020, Xiao2020}.
\item In the hydrodynamic regime the large negative coefficient is also due to the OMM as setting $\xi_1=0$ would result in a positive coefficient
\be
\tilde{\sigma}(q,\omega) \approx \varepsilon \,\omega_\mathrm{P}^2\frac{\Gamma_\mathrm{tot}}{q^2 v^2}\left[9.0+5.2\alpha^2\right].
\ee
\item In contrast, magnetoconductivity in the AIN regime is not affected by either the OMM or the second-order corrections and results solely from the leading-order term in the Berry curvature. The nonlocality of the quantum correction to the conductivity can be seen in its dependence on $q$, absent in the classical part.
\end{itemize}

\section{Skin effect}
\label{sec_skin}

\begin{figure} 	            
\includegraphics[width=0.8\columnwidth]{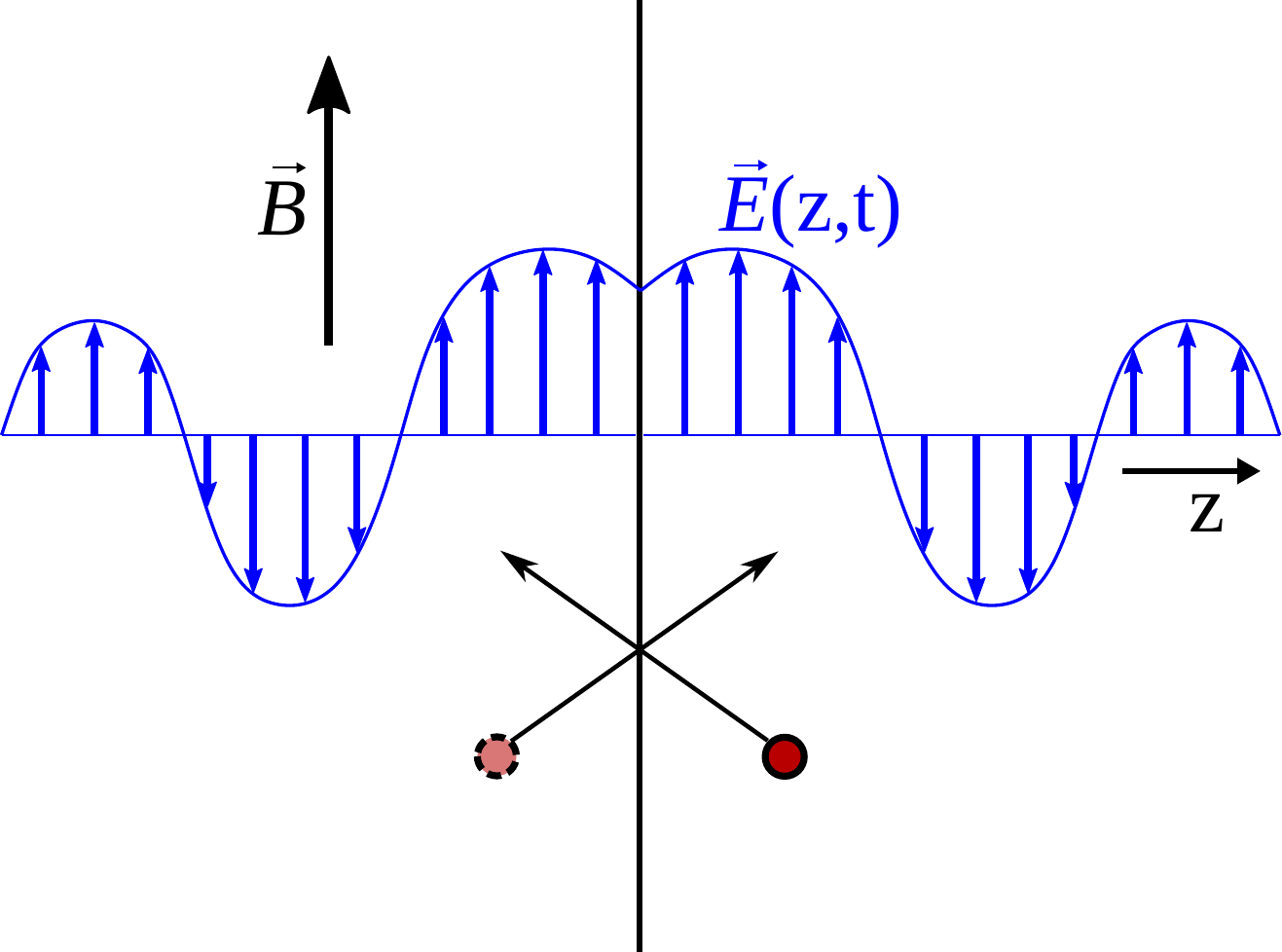}       
\caption{Implementation of specular boundary conditions. To the original system located in $z>0$ its mirror-reflected version is added in $z<0$, which creates a discontinuity in the derivative of the electric field at the boundary. A specular reflection of a particle can be seen as its mirror-reflected counterpart freely crossing the boundary.}
\label{fig_boundary}   
\end{figure} 

In order to find the skin depths and the surface impedance in the regimes identified in the previous section, we have to include the presence of a boundary in our analysis. This is done by imposing appropriate boundary conditions on the distribution functions. As a particularly simple choice we impose specular boundary conditions, meaning that during reflection the transverse components of velocity stay the same but the perpendicular component gets inverted; i.e., $\eta^{(s)}(z=0,v_x, v_y,v_z) = \eta^{(s)}(z=0,v_x, v_y, -v_z)$. These boundary conditions can be naturally implemented if, following ref. \cite{Lifshitz}, we add a mirror-symmetric counterpart (for $z<0$) to the original (physical) half-space system. Then, in the new setup the condition of specular reflection means that the particles move freely across the $z=0$ boundary (Fig. \ref{fig_boundary}). Because the electric field is parallel to the surface, its mirror-reflected components are related by $\vct E(-z,t) = \vct E(z,t)$. Furthermore, the external magnetic field $\vct B$ does not change under the mirror reflection, which allows us to use a single formula for the conductivity in the whole space.

From Maxwell's equations we obtain the relation (using $\vct E(z,t) = E(z,t)\vct{\hat x}$)
\be
\partial_z^2E(z, t) = \frac{1}{c^2}\partial_t^2E(z,t)+ \mu \partial_t J(z,t).
\label{maxwell_2}
\ee
Fourier-transforming the LHS:
\be
\begin{split}
& \int_{-\infty}^{\infty} \frac{dz}{2\pi} E''(z)e^{i q z} \\
& = \int_{-\infty}^{\infty}\frac{dz}{2\pi} \left(E'(z)e^{i q z}\right)'dz-i q\int_{-\infty}^{\infty} \frac{dz}{2\pi} E'(z)e^{i q z} dz.
\end{split}
\ee
The first derivative of $E(z)$ has a discontinuity at $z=0$, so that $E'(z=0^-)=-E'(z=0^+)$, which means that the integration of the first term should be done separately in $(-\infty,0)$ and $(0,\infty)$. The second term can be integrated by parts once again, producing
\be
\int_{-\infty}^{\infty} \frac{dz}{2\pi} E''(z)e^{i q z}= -\frac{1}{\pi}E'(z=+0)-q^2E(q).
\ee
The Fourier transform of $J(z,\omega)$ is $\sigma(q,\omega)E(q,\omega)$, where $\sigma(q,\omega)$ was determined in the previous section, so from Eq. (\ref{maxwell_2}) we obtain
\be
E(q,\omega) = \frac{-E'(z=0^+)}{\pi}\frac{1}{q^2 - \frac{\omega^2}{c^2} + i\mu\omega\sigma(|q|,\omega)},
\ee
where we explicitly write $|q|$ because conductivity has to be an even function of $q$. Finally, we obtain
\be
\begin{split}
E(z) & = -\frac{E'(z=0^+)}{\pi}\int_{-\infty}^{\infty}dq \frac{e^{-i q z}}{q^2 - \frac{\omega^2}{c^2} + i\mu\omega\sigma(|q|,\omega)} \\
& = -\frac{2E'(z=0^+)}{\pi}\int_{0}^{\infty}dq \frac{e^{-i q z}}{q^2 - \frac{\omega^2}{c^2} + i\mu\omega\sigma(q,\omega)}.
\end{split}
\label{integration_infinite}
\ee
This result allows us to calculate surface impedance, defined as \cite{Abrikosov}
\be
Z = \frac{E(z=0)}{\int_0^\infty J(z)dz}.
\ee
Using Maxwell's equations this formula can be rewritten as
\be
Z = -i\mu\omega\frac{E(z=0)}{E'(z=0^+)}.
\label{impedance}
\ee

As the conductivities in all the regimes except AIN show only quantitative changes with respect to their classical counterparts, we restrict ourselves to presenting the impedances obtained from Eq. (\ref{impedance}) and (\ref{integration_infinite}) in Table \ref{table_cond}.

\subsection*{Skin effect in the AIN regime}

\begin{figure} 	            
\includegraphics[width=1.0\columnwidth]{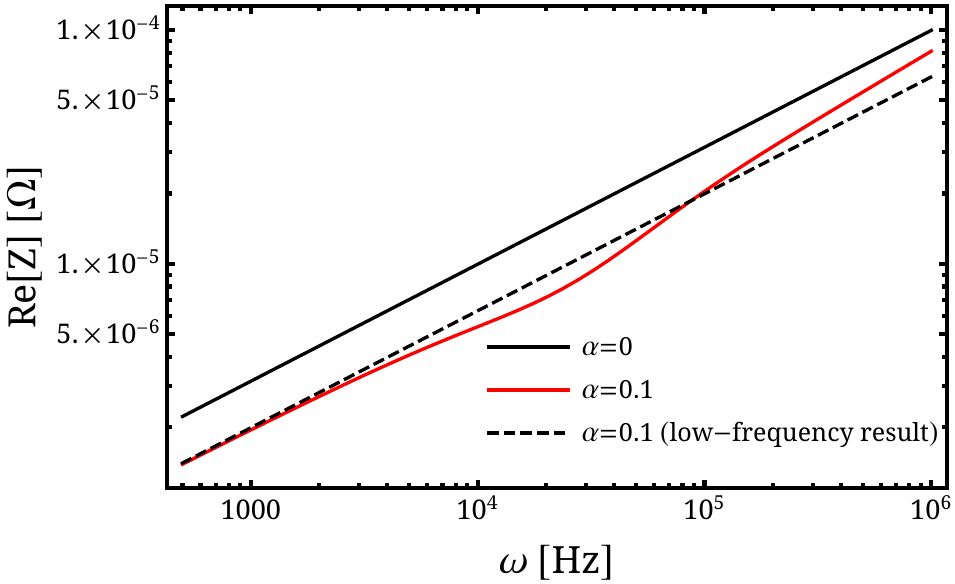}       
\caption{Dependence of the real part of the surface impedance $Z$ on frequency $\omega$ in the low-frequency normal and the AIN regimes, obtained from Eq. (\ref{impedance}, \ref{integration_infinite}, \ref{normal_cond_approx}) in the cases of no magnetic field ($\alpha=0$) and a non-zero magnetic field ($\alpha=0.1$). The dashed line is plotted on the basis of the impedance of the low-frequency normal regime as found in Table \ref{table_cond}, and it is the result that would be measured in the absence of the nonlocal behavior. The material parameters are the same as in section \ref{sec_conduct} i.e. $v=1.4\times10^5$ m/s, $\epsilon_\mathrm{F}=20$ meV, $\Gamma_\mathrm{mr}= 10^9$ Hz, $\Gamma_\mathrm{inter}=10^7$ Hz. The transition to the AIN regime takes place around $\omega = 10^5$ Hz. We see that changing the external magnetic field leads to a noticeable change of the surface impedance, especially shortly before the onset of the AIN regime, and moreover in this regime the scaling with frequency changes with respect to the no-field result.}
\label{fig_impedance}   
\end{figure}

We now present a detailed analysis of the skin effect in the most interesting case of the anomaly-induced nonlocal regime. In fact, we can also capture the crossover from the low-frequency normal regime by approximating the conductivity as
\be
\sigma(q,\omega) \approx \sigma_0\left[1 +\alpha^2\frac{\Gamma_\mathrm{mr}}{2\Gamma_\mathrm{inter}/3+q^2v^2/9\Gamma_\mathrm{mr}}\right],
\label{normal_cond_approx}
\ee
where $\sigma_0 = 2 \epsilon \omega_\mathrm{P}^2/\Gamma_\mathrm{mr}$. This approximation for the conductivity works only when $\Gamma_\mathrm{inter}\ll \Gamma_\mathrm{mr}$, $q v \ll \Gamma_\mathrm{mr}$ and $\omega\ll q^2v^2/\Gamma_\mathrm{mr}$. We want to evaluate the integral in Eq. ($\ref{integration_infinite}$), which can be done using standard contour integration methods. The resulting impedance in both the low-frequency normal and the AIN regimes can be seen in Fig. \ref{fig_impedance}.

The AIN regime can be observed when $(\omega\mu\sigma_0)v^2\gg\Gamma_\mathrm{mr}\Gamma_\mathrm{inter}$. For $z=0$ we integrate over the contour in the lower half-plane, and thus we are interested in the poles with a negative imaginary part. There are two such poles: $q_1 \approx e^{-i\pi/4}\sqrt{\mu\omega\sigma_0+i\alpha^2\frac{9\Gamma_\mathrm{mr}^2}{v^2}}$ and $q_2 \approx -i\sqrt{\Gamma_\mathrm{mr}\left(6\Gamma_\mathrm{inter}+9\alpha^2 \Gamma_\mathrm{mr}\right)}/v$. Please note that $|q_2|\ll|q_1|$. The electric field is
\be
\begin{split}
 E(z) = iE'(z=0^+)\left[\frac{\exp(-iq_1 z)}{q_1}+\alpha^2\frac{9\Gamma_\mathrm{mr}^2}{q_1^2 v^2}\frac{\exp(-iq_2 z)}{q_2}\right].
\end{split}
\label{mixed_electric}
\ee
This result shows that there is an anomalous component of the electric field with the skin depth much larger than the classical skin depth, so at large distances the anomalous component dominates over the classical one. The impedance is then
\be
Z = e^{i\pi/4}\sqrt{\frac{\mu\omega}{\sigma_0+i\alpha^2\frac{9\Gamma_\mathrm{mr}^2}{\mu\omega v^2}}}-\alpha^2\frac{9\Gamma_\mathrm{mr}}{\sigma_0 v}\left(6\frac{\Gamma_\mathrm{inter}}{\Gamma_\mathrm{mr}}+9\alpha^2\right)^{-\frac12}.
\label{mixed_impedance}
\ee
Therefore, in a finite magnetic field the impedance in this regime no longer scales with the square root of $\omega$, as seen in Fig. \ref{fig_impedance}. This regime with its characteristic impedance is a central finding of the present manuscript.

\section{Discussion and conclusions}

We provided a detailed analysis of the chiral magnetic conductivity across various transport regimes. In order to do so, we constructed a collision operator that captures three different relaxation mechanisms by exploiting its algebraic properties. We also included in our analysis second-order semiclassical corrections. The computed conductivities show a quantitative influence of these corrections in the hydrodynamic and anomalous regime, although it has to be noted that due to the coexistence of classical magnetoconductivity in these regimes, the quantum corrections could be difficult to separate out experimentally.

More importantly, we uncovered an entirely new transport regime where the conductivity is a combination of a classical local part and a nonlocal quantum part. This leads to a change of scaling of the surface impedance with the driving frequency when the material is placed in an external magnetic field. This phenomenon is related to a significant increase in the skin depth, which might be possible to observe in thin enough samples.

In the end, we note that there are various other phenomena that affect transport in Weyl semimetals that we did not include in our analysis. We neglected the influence of the Fermi arcs and the anomalous Hall effect (which plays a significant role in the propagation of light in time-reversal symmetry-breaking Weyl semimetals), as well as all the effects nonlinear in the electric field. Nonetheless, the results of this article hold for time-reversal symmetry-breaking Weyl semimetals placed in weak fields when the positions of the nodes are properly aligned with the electric field, as well as (qualitatively) for noncentrosymmetric Weyl semimetals. In particular, the presence of the anomaly-induced nonlocal regime should be discernible experimentally. 

Our work serves as a starting point for detailed analyses of transport regimes in Weyl semimetals, including different scattering mechanisms relevant for realistic experimental situations. As a result, one can construct a detailed theoretical description of a new generation of transport measurements beyond two-dimensional analogs. This allows one to investigate new effects absent in two dimensions, such as chiral vortical effect or anomalous thermal transport. It also permits to include realistic boundary conditions, important in electron hydrodynamics and usually neglected in high-energy applications of chiral kinetic theory.

\textit{Note added.} When this article was being prepared, a preprint appeared \cite{Sukhachov_skin_2021}, in which the existence of a nonlocal regime with the same qualitative features as the AIN regime was predicted. While \cite{Sukhachov_skin_2021}, contrary to the present work, assumes $\omega\gg \Gamma_{\mathrm{inter}}$, it is worth noting that when $\Gamma_{\mathrm{inter}}$ is replaced with $i\omega/2$ in Eq. (\ref{mixed_electric}), the results of the two articles are found to be in good agreement..

\section*{Acknowledgements}
We acknowledge discussions with Francisco Pe\~{n}a-Benitez, Peng Rao, and J\"{o}rg Schmalian. We thank Pavlo Sukhachov for comments on the manuscript. This work was in part supported by the Deutsche Forschungsgemeinschaft (DFG) through the Leibniz Program, the cluster of excellence ct.qmat (EXC 2147, project-id 39085490), as well as SFB 1143 (project-id 247310070). PS was also supported by the National Science Centre (NCN) Sonata Bis grant 2019/34/E/ST3/00405.
\appendix

\section{Higher order corrections to energy and Berry curvature}
\label{app_2nd}

Second-order corrections to the semiclassical equations of motion in the presence of the magnetic field were derived within the wave-packet approach in \cite{Niu2014, Niu2015}. These corrections modify the Berry curvature and the wave-packet energy, but do not otherwise change the form of the equations of motion [Eqs. (\ref{dotx})--(\ref{dotp})]. We present here formulas derived in Ref. \cite{Niu2015} and evaluate them for the Hamiltonian of a Weyl quasiparticle with chirality $s$,
\be
H(\vct p) = sv\bm{\sigma}\cdot\vct p.
\label{hamilt_appendix}
\ee
Let us denote the eigenstates $|u_n\rangle$ and eigenvalues $\epsilon_n$, with $n$ indicating the band. We label the upper band as $0$ and the lower band as $-1$. We define the interband Berry connection as $\vct A_{nm}=-i\langle u_n|\partial_{\vct p}|u_m\rangle$, and the velocity matrix as $\vct V_{nm}=\langle u_n|\partial_{\vct p}H(\vct p)|u_m\rangle$. We further define
\be
G_{nl} = -\frac{1}{2}\vct B\cdot\left(\Sigma_{m\neq l}\vct V_{nm}\times\vct A_{ml}+\vct V_{ll}\times\vct A_{nl}\right),
\ee
which is then used in the computation of the correction to the intraband Berry connection
\be
\vct a_0' = e\hbar\sum_{n\neq 0}\left[\frac{G_{0n}\vct A_{n0}}{\epsilon_0-\epsilon_n}\right]+e\hbar\frac{1}{4}\partial_{p_i}\left[\left(\vct B\times\vct A_{0n}\right)_i \vct A_{n0}\right]+\mathrm{c.c.}
\label{a0'}
\ee
Because in our model $\vct V_{-1-1}=-\vct V_{00}$, $G_{0-1}=G_{-10}=0$ and, consequently, the only contribution comes from the second term on the RHS of Eq. (\ref{a0'}). The resulting correction to the Berry curvature reads
\be
\vct \Omega' = \partial_{\vct p}\times\vct a_0' = - \frac{e \hbar \vct B}{4 |\vct p|^4} + \frac{e \hbar (\vct B\cdot \vct p)\vct p}{2 |\vct p|^6}.
\ee
This is taken into account in Eq. (\ref{berry}).

Second-order corrections to energy are composed of several terms [see Eq. (4) in \cite{Niu2015}]. Nonetheless, not all of them contribute to the final result. Firstly, there are two \textit{geometrical} terms: one of them depends on the Berry curvature and the magnetic moment
\be
\epsilon^{(1)} = \frac{e\hbar}{4}(\vct B\cdot\vct \Omega)(\vct B\cdot\vct m) = \frac{e^2\hbar^2v(\vct B\cdot \vct p)^2}{16|\vct p|^5},
\ee 
while the other one depends on the quantum metric $g_{ij} = \mathrm{Re}\langle\partial_{p_i}u_0|\partial_{p_j}u_0\rangle - A_{00,i}A_{00,j}$ and the inverse effective mass tensor $\alpha_{kl} = \partial_{p_k}\partial_{p_l}\epsilon_0$
\be
\epsilon^{(2)} =-\frac{e^2\hbar^2}{8}\epsilon_{sik}\epsilon_{tjl}B_sB_tg_{ij}\alpha_{kl} = -\frac{e^2\hbar^2v(\vct B\cdot \vct p)^2}{16|\vct p|^5},
\ee
and thus $\epsilon^{(1)} +\epsilon^{(2)} = 0$. All other second-order corrections to the energy derived in Ref. \cite{Niu2015} vanish identically for the Hamiltonian in Eq. (\ref{hamilt_appendix}), except for
\be
\epsilon^{(3)} =-e\hbar\vct B\cdot (\vct a_0'\times \vct V_{00}) = e^2\hbar^2v\frac{|\vct B|^2|\vct p|^2-(\vct B\cdot\vct p)^2}{8|\vct p|^5}.
\ee
The total energy including the second order corrections is presented in Eq. (\ref{energy}). We note that these corrections were also studied in \cite{Gorbar2017}.

\section{Eigenfunctions of the collision operator}
\label{app_eigenfunctions}

The inner product of Eq. (\ref{inner}) can be rewritten in the spherical coordinates $(p, \theta, \phi)$ in the momentum space as
\be
\begin{split}
& \langle \eta |\zeta \rangle  = \int \frac{d^3 p}{(2\pi\hbar)^3} D(\vct{p})\delta\left[\epsilon_\mathrm{M}(\vct p) - \epsilon_\mathrm{F}\right]\eta(\vct p)^* \zeta(\vct p) \\
& = \int \frac{d(\cos\theta)d\phi}{(2\pi\hbar)^3} D(\vct{\hat p}) \frac{p_\mathrm{F}(\vct{\hat{p}})^2}{v_\mathrm{M}(\vct{\hat{p}})}\eta(\vct{\hat{p}})^* \zeta(\vct{\hat{p}}),
\end{split}
\label{inner_new}
\ee
where $v_\mathrm{M}(\vct{\hat p}) = \vct v_\mathrm{M}(\vct{\hat p})\cdot\vct{\hat p}$. Hereafter, we suppress the $(s)$ labels on all quantities. We furthermore define
\be
\mathcal{V}(\vct{\hat{p}}) = D(\vct{\hat p}) \frac{p_\mathrm{F}(\vct{\hat{p}})^2}{v_\mathrm{M}(\vct{\hat{p}})}.
\ee

In the absence of spherical symmetry, we can find physically sensible expressions for the eigenfunctions of the single-species collision operator $\hat{C}$ defined below Eq. (\ref{collision_general}) by exploiting its algebraic properties and a few facts:

(1) The eigenfunctions, which we denote $K_l^m$, should be perturbations of spherical harmonics
\be
K_l^m = Y_l^m + O(\alpha).
\ee

(2) The eigenfunctions are orthogonal with respect to the new inner product, which follows from the hermiticity of $\hat{C}[\eta]$
\be
\int \frac{d^2 \hat{p}}{(2\pi\hbar)^3}\mathcal{V}(\vct{\hat{p}}) K_{l'}^{m'}(\hat{\vct p})^* K_{l}^{m}(\hat{\vct p}) = \delta_{l, l'}\delta_{m, m'}.
\label{orthogonality}
\ee

(3) Let us expand the distribution function as $\eta = \sum_{l,m}X_l^mK_l^m(\vct{\hat{p}})$, where $X_l^m = X_l^m(\vct x, t)$. The rate of particle number change is
\begin{multline}
\frac{d \langle n \rangle}{d t } = \\
\int \frac{d^3 p}{(2 \pi \hbar)^3} D(\vct{\hat p}) \frac{d f}{d t} = -\int \frac{d^3 p}{(2 \pi\hbar)^3} D(\vct{\hat p}) \delta\left[\epsilon_\mathrm{M}(\vct p) - \epsilon_\mathrm{F}\right]\hat{C}[\eta ]   \\ 
= \sum_{l,m} \int \frac{d^2 \hat{p}}{(2 \pi\hbar)^3}\mathcal{V}(\vct{\hat{p}}) \Gamma_{l,m} X^m_l(\vct x) K^m_l(\vct{\hat p}).
\end{multline}
For processes that conserve the particle number, there must be an eigenvector of $\hat{C}$ associated with the eigenvalue $0$ and corresponding to the particle number. Because we are considering a perturbation to the no-field problem, in which the corresponding eigenfunction was $Y_0^0$, now the corresponding eigenfunction has to be $K_0^0$ and the eigenvalue associated to it is $\Gamma_{0,0} = 0$. Therefore, the RHS of the above equation is zero when
\be
\int \frac{d^2 \hat{p}}{(2 \pi\hbar)^3} \mathcal{V}(\vct{\hat{p}}) K^m_l(\vct{\hat p})=0,~~~~~~\forall l \geq 1
\ee
which, when taking Eq. (\ref{orthogonality}) into account, implies that $K_0^0(\hat{\vct p}) \propto 1 \propto Y_0^0(\hat{\vct p})$.

(4) The rate of momentum change is
\begin{multline}
\frac{d \langle \vct p \rangle}{d t } = \int \frac{d^3 p}{(2 \pi\hbar)^3}D(\vct{\hat p})\vct p \frac{d f}{d t} \\
\propto -\int \frac{d^3 p}{(2 \pi\hbar)^3} D(\vct{\hat p}) \delta\left[\epsilon_\mathrm{M}(\vct p) - \epsilon_\mathrm{F}\right] p Y_1^{M}(\hat{\vct p})\hat{C}[\eta ]   \\ 
= \frac{\epsilon_\mathrm{F}}{v}\sum_{l,m} \int \frac{d^2 \hat{p}}{(2 \pi\hbar)^3} \mathcal{V}(\vct{\hat{p}}) \left\{\frac{v p_\mathrm{F}(\vct{\hat{p}})}{\epsilon_\mathrm{F}} Y_1^{M}(\hat{\vct p})\right\}\\
\times \Gamma_{l,m} X^m_lK^m_l(\vct{\hat p}),
\end{multline}
where $M = -1, 0, +1$ for the different momentum components. A reasoning similar as in paragraph (3) dictates that for processes that conserve momentum and the particle number $\Gamma_{1,M}=0$ and 
\be
\int \frac{d^3 p}{(2 \pi\hbar)^3}\left\{\frac{v p_\mathrm{F}(\vct{\hat{p}})}{\epsilon_\mathrm{F}} Y_1^{M}(\hat{\vct p})\right\}\mathcal{V}(\vct{\hat{p}})   K^m_l(\vct{\hat p})=0,~~~~\forall l \geq 2
\ee
which, when taking Eq. (\ref{orthogonality}) into account, implies that
\begin{multline}
\frac{v p_\mathrm{F}(\vct{\hat{p}})}{\epsilon_\mathrm{F}} Y_1^{M}(\hat{\vct p}) = \sum \lambda^M_{M'} K_1^{M'}(\hat{\vct p})+ \mu^M K_0^0(\hat{\vct p}) \\
\propto \sum \lambda^M_{M'} K_1^{M'}(\hat{\vct p})+ \nu^M Y_0^0(\hat{\vct p}),
\end{multline}
where $\lambda^M_{M'}$, $\mu^M$, $\nu^M$ are some coefficients. It follows that $K_1^M$ is a linear combination of $p_\mathrm{F}(\vct{\hat{p}}) Y_1^{M'}(\hat{\vct p})$ and $Y_0^0$, with coefficients chosen such that the orthogonality relation is satisfied. (The matrix $\lambda^M_{M'}$ is of the form $1-O(\alpha)$, $1$ being the identity matrix, and therefore it is invertible.)

We take all modes with $L\geq 2$ to be eigenfunctions to the same eigenvalue. Hence, all modes with $L\geq 2$ constitute one eigenspace orthogonal to the eigenvectors with $L=0, 1$. We can choose the basis of this eigenspace arbitrarily, as long as it satisfies the orthogonality condition [Eq. (\ref{orthogonality})].

Let us orthonormalise the ordered set of basis vectors $\left(Y_0^0(\hat{\vct p}), \frac{v p_\mathrm{F}(\vct{\hat{p}})}{\epsilon_\mathrm{F}}Y_1^0(\hat{\vct p}), \frac{v p_\mathrm{F}(\vct{\hat{p}})}{\epsilon_\mathrm{F}}Y_1^1(\hat{\vct p}), \frac{v p_\mathrm{F}(\vct{\hat{p}})}{\epsilon_\mathrm{F}}Y_1^{-1}(\hat{\vct p})\right)$ with regard to the inner product in Eq. (\ref{inner_new}) using the Gram-Schmidt process. This gives the vectors $K_0^0(\hat{\vct p})$, $K_1^0(\hat{\vct p})$, $K_1^1(\hat{\vct p})$, $K_1^{-1}(\hat{\vct p})$. The Gram-Schmidt process guarantees that $K_0^0\propto Y_0^0$, which satisfies condition (3), and that $K_1^M$ are combinations of $p_\mathrm{F}(\vct{\hat{p}}) Y_1^{M'}$ and $Y_0^0$, which satisfies condition (4). The formulas for $K_1^M$ depend on the order in which the basis vectors are orthonormalized, but the space spanned by them does not (as it is defined as the 3D subspace orthogonal to $Y_0^0$ in the space spanned by $\{Y_0^0, \frac{v p_\mathrm{F}(\vct{\hat{p}})}{\epsilon_\mathrm{F}}Y_1^0, \frac{v p_\mathrm{F}(\vct{\hat{p}})}{\epsilon_\mathrm{F}}Y_1^1, \frac{v p_\mathrm{F}(\vct{\hat{p}})}{\epsilon_\mathrm{F}}Y_1^{-1}\}$). Consequently, we can unambiguously define the projection operators
\be
P_0 = |K_0^0\rangle\langle K_0^0|,
\ee
\be
P_1 = \sum_{M=-1,0,1}|K_1^M\rangle\langle K_1^M|,
\ee
\be
P_{\mathrm{higher}} = 1 - P_0 - P_1.
\ee
The eigenfunctions can be expanded to the appropriate order, e.g. up to $\alpha^2$. Using the definition of the inner product in Eq. (\ref{inner_new}), it is possible to evaluate the action of the projection operators on any distribution function. In the main text we consider two coupled particle species that exchange particles, so that the $(s)$ labels appear and the projections operators are defined as in Eq. (\ref{projectors}).

\section{Perturbative solution of the Boltzmann equation}
\label{app_solution}

\begin{table*}[]
\begin{tabular}{|l|c|c|}
\hline
Regime & Nonzero $B_L^M$ coefficients up to $L=2$ & $C_1^{\pm1}$ coefficients \\
\hline 
Low-frequency normal& 
\begin{tabular} {@{}c@{}c@{}} $B_0^0 = -e v E\sqrt{\pi}\left[\frac{1}{\Gamma_\mathrm{inter}}-\frac{4}{3\Gamma_\mathrm{mr}}\right]$ \\ $B_2^0 = e v E\sqrt{\frac{\pi}{5}}\frac{2}{3\Gamma_\mathrm{mr}}$ \\ $B_2^{\pm 2} = -e v E\sqrt{\frac{2\pi}{15}}\frac{1}{\Gamma_\mathrm{mr}}$ \end{tabular} &
 $C_1^{\pm 1} = \mp e v E\sqrt{\frac{2\pi}{3}}\frac{1}{\Gamma_\mathrm{mr}}$  \\ 
\hline
Anomaly-induced nonlocal & 
\begin{tabular} {@{}c@{}c@{}c@{}} $B_0^0 = -e v E\sqrt{\pi}\frac{6 \Gamma_\mathrm{mr}}{q^2v^2}$ \\ $B_1^0 = -ievE\sqrt{3\pi}\frac{2}{qv}$ \\$B_2^0 = \sqrt{\frac{\pi}{5}} e v E\left[\frac{4}{\Gamma_\mathrm{tot}}+\frac{2}{3\Gamma_\mathrm{mr}}\right]$ \\ $B_2^{\pm 2} = -e v E\sqrt{\frac{2\pi}{15}}\frac{1}{\Gamma_\mathrm{mr}}$ \end{tabular} &
 $C_1^{\pm 1} = \mp e v E\sqrt{\frac{2\pi}{3}}\frac{2\left(3\Gamma_\mathrm{mr}+\Gamma_\mathrm{tot}\right)^2}{15 \Gamma_\mathrm{mr}^2 \Gamma_\mathrm{tot}}$ \\ 
\hline
Hydrodynamic & 
\begin{tabular} {@{}c@{}c@{}} $B_0^0 = e v E\frac{11.3\Gamma_\mathrm{tot}}{q^2 v^2}$ \\ $B_2^0 = e v E\frac{2.4\Gamma_\mathrm{tot}}{q^2 v^2}$ \\ $B_2^{\pm 2} = -e v E\sqrt{\pi}\frac{2.9\Gamma_\mathrm{tot}}{q^2 v^2}$ \end{tabular} &
 $C_1^{\pm 1} = \mp e v E\frac{10.2\Gamma_\mathrm{tot}}{q^2 v^2}$ \\ 
\hline
Ballistic & 
 $B_0^0 = -e v E\frac{0.2}{q v}$&
 $C_1^{\pm 1} = \pm e v E\frac{3.0}{q v}$ \\
\hline
High-frequency normal & 
\begin{tabular} {@{}c@{}c@{}} $B_0^0 = ie v E\sqrt{\pi}\frac{2}{3\omega}$ \\ $B_2^0 = -ie v E\sqrt{\frac{\pi}{5}}\frac{2}{3\omega}$ \\ $B_2^{\pm 2} = ie v E\sqrt{\frac{2\pi}{15}}\frac{1}{\omega}$\end{tabular} &
 $C_1^{\pm 1} = \pm i e v E\sqrt{\frac{2\pi}{3}}\frac{1}{\omega}$ \\
\hline
\end{tabular}
\caption{Coefficients $B_L^M$ with low $L$ and $C_1^{\pm 1}$ obtained from the perturbative solution of the Boltzmann equation. Only the leading-order contributions in the respective regimes are shown. The numerical coefficients in the hydrodynamic and ballistic regimes are rounded up to the first decimal place.}
\label{table_solutions}
\end{table*}

Before presenting the full perturbative solution, let us first explain a subtlety in the evaluation of the collision integral. The operator $P_0^{(s)}$ in Eq. (\ref{projectors}) acts on both distribution functions $|\eta^{(\pm s)}\rangle$, therefore coupling the systems of equations for $(s)$ and $(-s)$, which would suggest that the number of equations that need to be solved has doubled. However, we can use the following observation: the factor of $s$ in the equations of motion always appears in the combination $s\,\alpha$ (see the formulas in sections \ref{sec_kinetic}, \ref{sec_linearize}). Therefore, in the expansions of $\langle K_0^{0,(s)}|\eta^{(s)}\rangle$ and $\langle K_0^{0,(-s)}|\eta^{(-s)}\rangle$ in $\alpha$, the even-order terms have to be equal, and the odd-order terms have to be opposite. This in turn means that, as we are interested in the expansion up to the second order in $\alpha$,
\be
P_0^{(s)}[\eta^{(s)},\eta^{(-s)}] = 2|K_0^{0,(s)}\rangle\left\{\langle K_0^{0,(s)}|\eta^{(s)}\rangle\right\}_{\alpha}
\label{P0trick}
\ee
where $\{\ldots\}_{\alpha}$ denotes evaluating only the linear order in $\alpha$. This way the doubling of the number of equations that need to be solved is avoided.

The solution at the 0-th order in $\alpha$ was presented in section \ref{sec_conduct}, except for one subtlety. The actual form of Eq. (\ref{aux2}) reads
\begin{multline}
i\omega A_L^M -i q v \sqrt{\frac{(2L-1) }{(2L+1)}}C_{10,(L-1)0}^{L0} C_{10,(L-1)M}^{LM} A_{L-1}^M \\ -iq v \sqrt{\frac{(2L+3) }{(2L+1)}}C_{10,(L+1)0}^{L0} C_{10,(L+1)M}^{LM} A_{L+1}^M = -\Gamma_\mathrm{tot} A_L^M.
\label{auxLM}
\end{multline}
The Clebsch-Gordan coefficients can be expressed explicitly to obtain
\be
\sqrt{\frac{(2L-1) }{(2L+1)}}C_{10,(L-1)0}^{L0} C_{10,(L-1)M}^{LM} = \sqrt{\frac{L^2-M^2}{4L^2 - 1}},
\ee
\be
\sqrt{\frac{(2L+3) }{(2L+1)}}C_{10,(L+1)0}^{L0} C_{10,(L+1)M}^{LM} = \sqrt{\frac{(L+1)^2-M^2}{4(L+1)^2 - 1}},
\ee
so that for $L\gg M$ the coefficients above all tend to a constant value $1/2$. In fact, already when taking $L=2$, $M=1$, the exact values $\sqrt{3/15}\approx 0.45$ and $\sqrt{8/35}\approx 0.48$ are very close to $1/2$, so approximating these coefficients as $1/2$ in Eq. (\ref{aux2}) does not produce a noticeable error.

Let us turn our attention to the system of equations at the 1st order in $\alpha$. The exact forms of the equations at this order are too lengthy to present. Let us however make a few remarks. At linear order in $\alpha$, the electric field enters the equations for $L=0$, $M=0$ and $L=2$, $M=-2,0,2$. The equation for $L=0$, $M=0$ contains a term proportional to $\Gamma_\mathrm{inter}$ in agreement with Eq. (\ref{P0trick}). The equations for $L\geq 3$ asymptotically tend to the recurrence relation which at this level reads
\begin{multline}
\left(i\omega+\Gamma_\mathrm{tot}\right) B_L^M -i q v \frac{1}{2} \left[B_{L-1}^M + B_{L+1}^M\right] \\
+\frac14r^{L-2}(1-r^2)\left(A_1^{M-1}- A_1^{M+1}\right)= 0
\end{multline}
and its general solution is
\be
B_L^M = B_2^M r^{L-2}+X^M (L-2)r^{L-3},
\ee
where
\be
X^M = -\frac14(r+r^3)\left(A_1^{M-1}- A_1^{M+1}\right).
\ee
These formulas are then plugged into the equations for $L=0,1,2$, and the system of equations is solved. Because the exact solution is too complicated, in Table \ref{table_solutions} we present only the formulas obtained in the limiting cases corresponding to the different transport regimes, showing the leading-order terms and neglecting the coefficients that evaluate to zero.

The solutions at the 0-th and 1-st orders are then used to find the solutions at the 2nd order in $\alpha$. Here the electric field enters equations for $L=1$, $M=\pm 1$ and $L=3$, $M=\pm1, \pm3$. Again, the exact forms of these equations are too lengthy to present. The asymptotic form of the equations for $L\geq 4$ can be written as the recurrence relation
\begin{multline}
8r^4\left[(1+r^2)C_L^M-rC_{L-1}^M-rC_{L+1}^M\right]\\
+r^L\left[-13+10r^2-5r^4+2L-2 Lr^4\right]\left(X^{M-1}-X^{M+1}\right) \\
+2r^{L+1}(1-r^4)\left(B_2^{M-1}-B_2^{M+1}\right)+ 4r^{L+3}(1+r^2)A_1^M=0.
\end{multline}
The general solution is
\be
C_L^M = C_3^Mr^{L-3}+Y^M(L-3)r^{L-4}+Z^M(L-3)^2r^{L-4},
\ee
where
\begin{multline}
Y^M = \frac{(3-6r^2+5r^4)\left(X^{M-1}-X^{M+1}\right)}{4(1-r^2)}\\
-\frac{r(1-r^4)\left(B_2^{M-1}-B_2^{M+1}\right)+2r^3(1+r^2)A_1^M}{4(1-r^2)}
\end{multline}
and
\be
Z^M=-\frac18(1+r^2)\left(X^{M-1}- X^{M+1}\right).
\ee
These formulas are then plugged into the equations for $L=0,1,2,3$, and the system of equations is solved. Because at this level of approximation only the $C_1^{\pm 1}$ components contribute to the final expression for the current, in Table \ref{table_solutions} we present only the formulas for $C_1^{\pm 1}$ in the different transport regimes.

\section{The classical Lorentz force and the magnetization current}
\label{app_estimate}

\begin{figure} 	            
\includegraphics[width=0.5\columnwidth]{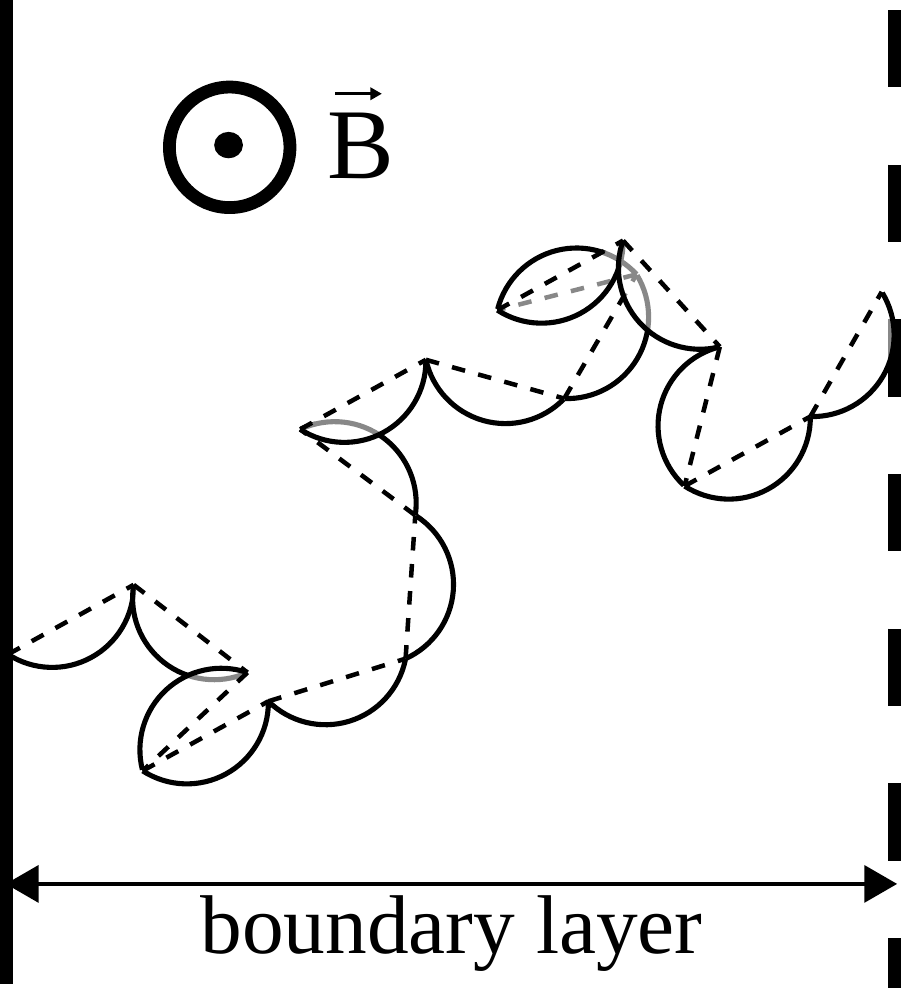}       
\caption{The effect of the Lorentz force on transport in the hydrodynamic regime. The loss of momentum takes place due to the particle performing a random walk across the boundary layer. The full path traversed by a particle in the presence of magnetic field (solid line) is longer than in its absence (dashed line).}
\label{fig_random_walk}   
\end{figure} 

In the main text we neglected the the classical Lorentz force, which shows up as the last term on the RHS of Eq. (\ref{momentum_term}) and contributes $e v\left[\vct{\hat p} \times\vct B\right]\cdot\partial_{\vct p}\eta^{(s)}$ to the linearized Boltzmann equation at the classical level. Taking $\vct B = B\vct{\hat x}$, this term evaluates to:
\be
e v B \left(\hat{p}_z \partial_{p_y}-\hat{p}_y \partial_{p_z}\right)\eta^{(s)}.
\label{lorentz}
\ee
This term does not change the conductivity in either the normal regimes or in the AIN regime when $\vct E$ is parallel to $\vct B$. The reason is that in all of these regimes, the distribution function at the classical level is proportional to $E\,\hat{p}_x$, and evaluating Eq. (\ref{lorentz}) gives 0. 

The situation is different, however, in the hydrodynamic and anomalous regimes. Let us start with the anomalous regime, where we can calculate the change of conductivity perturbatively in $B$. The linearized Boltzmann equation at the classical level is (neglecting the label $(s)$):
\begin{multline}
\left(i\omega+\frac{1}{\Gamma}\right)\eta - i q v \hat{p}_z\eta + e v E \hat{p}_x \\
+ e v B \left(\hat{p}_z \partial_{p_y}-\hat{p}_y \partial_{p_z}\right)\eta = 0.
\end{multline}
Here $\Gamma$ is some relaxation time; it does not matter which one, as all relaxation times, as well as $\omega$, are much smaller than $q v$. The leading order solution in the anomalous regime is 
\be
\eta^0 = -i\frac{e v E\hat{p}_x}{q v \left(\hat{p}_z+i\delta\right)}
\ee
with $\delta$ infinitesimal. Solving perturbatively in $B$ at the next two orders produces
\begin{align}
\eta^1&=-\frac{e^2vEB }{q^2\epsilon_\mathrm{F}}\frac{\hat{p}_x\hat{p}_y}{\hat{p}_z^3 + i\delta},\\
\eta^2&=i\frac{e^3v^2EB^2}{q^3\epsilon_\mathrm{F}^2}\left(\frac{\hat{p}_x}{\hat{p}_z^3 + i\delta}+\frac{3\hat{p}_x\hat{p}_y^2}{\hat{p}_z^5 + i\delta}\right).
\end{align}
Calculating the current and multiplying by 2 due to the two nodes $(s)$ and $(-s)$ gives 
\begin{multline}
\vct J = 2e\int\frac{d^3 p}{(2\pi\hbar)^3}v\vct{\hat p}\partial_{\epsilon_\mathrm{M}}f_0 \eta \\
= \left(1+\frac{e^2 v^2 B^2}{4q^2\epsilon_\mathrm{F}^2}\right)\frac{3\pi}{2}\frac{\epsilon\omega_\mathrm{P}^2}{q v}\vct E.
\end{multline}
So, magnetic field causes positive magnetoconductivity. Using the parameters for WP$_2$ \cite{Kumar2017, Gooth2018} and taking $q v = 5 \Gamma_\mathrm{tot}$, we obtain $e^2 v^2/q^2\epsilon_\mathrm{F}^2 \approx 10^{-3}$ T$^{-2}$, while for comparison the magnitude of quantum effects is $\alpha^2/|\vct B|^2 \approx e^2 \hbar^2 v^4/ \epsilon_\mathrm{F}^4 \approx 10^{-8}$ T$^{-2}$.

In the hydrodynamic regime, we estimate the classical magnetoconductivity by resorting to the following picture. The relaxation of momentum happens, when particles traverse the skin layer while performing a random walk with the step size $l_{\mathrm{mc}}$. The presence of the magnetic field curves the path and makes the total path traversed between the two boundaries of the skin layer longer: see Fig. \ref{fig_random_walk}. For a particle whose $x$ component of the momentum reads $\epsilon_\mathrm{F}\hat{p_x}/v$, the radius of the cyclotron orbit is $r_c = \frac{\epsilon_\mathrm{F}}{e v B\sqrt{1-\hat{p}_x^2}}$. The length of the arc traversed by a particle between two collisions is roughly equal to $l_{\mathrm{mc}}\left(1+\frac{l_{\mathrm{mc}}^2}{8r_c^2}\right)$. Because conductivity is proportional to the relaxation time, we can hypothesise that the effective $\Gamma_\mathrm{tot}$ changes to $K(\hat{\vct p}) \Gamma_\mathrm{tot}$ where
\be
K(\hat{\vct p}) = \left(1+\frac{e^2v^4B^2(1-\hat{p}_x^2)}{\Gamma_\mathrm{tot}^2\epsilon_\mathrm{F}^2}\right).
\ee
Then, the total current is of the order
\be
\begin{split}
\vct J &\approx \int\frac{d^3 p}{(2\pi\hbar)^3}\vct{\hat p}\partial_{\epsilon_\mathrm{M}}f_0 \frac{e^2 v^2 E \Gamma_\mathrm{tot}\hat{p}_x}{q^2v^2}K(\hat{\vct p}) \\
&\approx \left(1+\frac{e^2 v^4 B^2}{\Gamma_\mathrm{tot}^2\epsilon_\mathrm{F}^2}\right)\frac{\Gamma_\mathrm{tot}}{q^2 v^2}\epsilon\omega_\mathrm{P}^2\vct E.
\end{split}
\ee
So in this regime the classical magnetoconductivity is also positive. Using the parameters for WP$_2$ \cite{Kumar2017, Gooth2018}, we obtain $e^2 v^4/\Gamma_\mathrm{tot}^2\epsilon_\mathrm{F}^2 \approx 10^{-2}$ T$^{-2}$, while for comparison the magnitude of quantum effects is $\alpha^2/|\vct B|^2 \approx e^2 \hbar^2 v^4/\epsilon_\mathrm{F}^4 \approx 10^{-8}$ T$^{-2}$.

Let us now show that the magnetization current [Eq. (\ref{J_magn})] is zero. It can be written as
\be
\vct J_{\mathrm{magn}} = \nabla\times\vct M,
\ee
where
\be
\vct M=\sum_{s=\pm 1} \int \frac{d^3 p}{(2\pi)^3}D(\vct p)^{(s)}\frac{s e \hbar v}{2 |\vct p|} \hat{\vct p} f^{(s)}(q,\vct p)
\label{magnetization}
\ee
is the total magnetization. Because the distribution functions only change along $z$, we have 
\be
\vct J_{\mathrm{magn}} = -iq\left(\vct{\hat{y}}M_x-\vct{\hat{x}}M_y\right).
\ee
Due to the symmetry of the problem, $M_y=0$. However, also the total magnetization in the $\hat{x}$ direction is identically null, which can be inferred as follows. The expression (\ref{magnetization}) can be expanded in $\alpha$ in the same way as the current in Eq. (\ref{J_expanded}):
\be
\begin{split}
& M_x   = -\frac{e \epsilon_\mathrm{F}}{v(2\pi\hbar)^3}\times\\
&\sum_{s = \pm 1}s\int d(\cos\theta)d\phi  \left[1+s\alpha \hat{p}_x\right]\hat{p}_x\eta^{(s)}(q, \vct{\hat p}).
\end{split}
\label{M_expanded}
\ee
The sum over $s$ selects contributions on the order $\alpha$. There are two such contributions. One comes from evaluating the integral of $\alpha \hat{p}_x^2\left\{\eta^{(s)}\right\}_{\alpha=0}$, where $\left\{\eta^{(s)}\right\}_{\alpha=0}$ is the distribution function at the classical order, and the result of the integration is a combination of $A_0^0$, $A_2^0$ and $A_2^{\pm 2}$ (which correspond to modes even in $\hat{p}_x$). However, these coefficients are zero as seen in Eq. (\ref{classical_solution}). The second contribution comes from evaluating the integral of $\hat{p}_x\left\{\eta^{(s)}\right\}_{\alpha}$, where $\left\{\eta^{(s)}\right\}_{\alpha}$ is the distribution function at the linear order in $\alpha$, and the result of the integration is a combination of $B_1^{\pm1}$ (which correspond to modes odd in $\hat{p}_x$). These coefficients, however, are also zero, as seen in Table \ref{table_solutions}. Therefore, the total $M_x$ summed over the two valleys is zero, and thus $\vct J_{\mathrm{magn}}=0$.

\bibliography{Skin_Effect} 

\end{document}